\documentclass[fleqn,usenatbib]{mnras}

\usepackage[T1]{fontenc}
\usepackage{ae,aecompl}
\usepackage{multirow}

\usepackage{graphicx}	
\usepackage{amsmath}	
\usepackage{amssymb}	
\usepackage{comment}
\usepackage{bm}
\usepackage{ulem}

\def\ee{\end{equation}}
\def\be{\begin{equation}}
\def\GW{{\rm GW}}

\def\gpcyr{\rm \,Gpc^{-3}\,yr^{-1}}
\def\Rmrgb{R_{\bullet\bullet,\rm B}}
\def\Rmrgd{R_{\bullet\bullet,\rm D}}
\def\Rmrgns{R_{\rm BNS}}
\def\Rmrg{\bm{R}_{\rm mrg}}
\def\mc{{{\mathcal{M}}}_{\rm c}}

\title[GW background and eccentric compact binaries]{Stochastic Gravitational Wave Background 
and Eccentric Stellar Compact Binaries 
}
\author[Zhao \& Lu]{Yuetong Zhao$^{1,2}$ and Youjun Lu$^{1,2,\dagger}$ \\
$^{1}$\,CAS Key Laboratory for Computational Astrophysics, National Astronomical Observatories, Chinese Academy of Sciences, \\ 20A Datun Road, Beijing 100101, China; $^\dagger$\,luyj@nao.cas.cn\\
$^{2}$\,School of Astronomy and Space Sciences, University of Chinese
Academy of Sciences, 19A Yuquan Road, Beijing 100049, China
}

\date{Accepted XXX. Received YYY; in original form ZZZ}

\hypersetup{draft}

\begin{document}
\label{firstpage}
\pagerange{\pageref{firstpage}--\pageref{lastpage}}
\maketitle
%
\begin{abstract}
Gravitational wave (GW) radiations from numerous cosmic stellar-compact-binaries form a stochastic GW background (GWB), which is expected to be detected by ground and space GW detectors in future. Theoretical predictions of this GWB were mostly obtained by assuming either circular binaries and/or a specific channel for binary formation, which may have some uncertainties. In this paper, we estimate the GWB and its spectrum by using simple models for the formation of both stellar mass binary black holes (sBBHs) and binary neutron stars (BNSs). We consider that the dynamically originated sBBHs have relatively larger masses and higher eccentricities compared with those from field binary stars and its possible effect on the GWB spectrum. We find that the GWB spectrum may have a turnover in the low-frequency (Laser Interferometer Space Antenna; LISA) band and may be better described by a broken double power-law than a single power-law with the unique index $2/3$, and the low-frequency slope depends on the significance of the dynamically originated sBBHs with high eccentricities. We further generate mock samples of sBBHs and BNSs, and simulate the mock GWB strain in the time domain. We find that GWB can be detected with signal-to-noise ratio (SNR) $\gtrsim274/255/21$ by LISA/Taiji/TianQin over $5$-years' observation and $\gtrsim3$ by LIGO over $2$-years' observation. 
Furthermore, we estimate that the number of sBBHs that may be detected by LISA/Taiji/TianQin individually with SNR $\gtrsim8$ is $\sim5$-$221$/$7$-$365$/$3$-$223$ over $5$-years' observation. 
\end{abstract}



\section{Introduction}
\label{sec:intro}

Gravitational waves (GWs) were recently directly detected by the advanced Laser Interferometer Gravitational wave Observatories (LIGO), which announced the new era of gravitational wave astronomy \citep{2016PhRvL.116f1102A}. During the O1 and O2 operations of LIGO and VIRGO \citep{2015CQGra..32b4001A}, at least ten mergers of stellar binary black holes (sBBHs) and one merger of binary neutron stars (BNSs) were detected at high frequencies ($\sim 10-10^3$\,Hz; e.g., \citealt{2016PhRvL.116f1102A, 2016PhRvL.116x1103A, 2017PhRvL.118v1101A, 2017ApJ...851L..35A, 2017PhRvL.119n1101A, 2017PhRvL.119p1101A, 2019PhRvX...9c1040A}). Such mergers of sBBHs and BNSs are now regularly being detected by the LIGO/VIRGO O3 operation \citep[e.g.,][https://gracedb.ligo.org]{2018LRR....21....3A}. KAGRA \citep{2012CQGra..29l4007S, 2013PhRvD..88d3007A} came on line on 25 February 2020 but couldn't join the O3 operation because O3 finishes earlier than planned. GWs emitted from individual sBBHs at their inspiral stage, long before the final merger stage, will be detected by future space GW detectors [Laser Interferometer Space Antenna, LISA, \url{https://lisa.nasa.gov}; Taiji \citep{2017SSPMA..47a0404H, 2019arXiv190907104R}; and TianQin \citep{2016CQGra..33c5010L, 2019PhRvD.100d3003W}] at lower frequencies ($\sim 10^{-4}-1$\,Hz), though the data analysis will be a great challenge \citep[e.g.,][]{2018APS..APRH14001B, 2019AAS...23314108B}. Multi-band GW observations will be possible in future by combining space and ground-based GW detectors together \citep{2016PhRvL.116w1102S}. It is also expected that LIGO/Virgo/KAGRA and LISA will detect the stochastic background composed of GWs emitted from the merger and inspiral of cosmic sBBHs and BNSs at both high and low frequencies \citep[e.g.,][]{2012CQGra..29l4016A, 2016PhRvL.116w1102S, 2017ogw..book...43C}. 

How many mergers of sBBHs and BNSs can be detected and how large the low frequency GW background (GWB) depends on the merger rates and detailed properties of these systems, and thus their actual formation mechanisms \citep[see][]{2018arXiv180910360C, 2018MNRAS.481.4775D}, which are still not clear and interesting to investigate. The formation mechanisms of sBBHs and BNSs have been extensively studied \citep[e.g.,][]{2002ApJ...572..407B, 2007PhR...442...75K}, especially after the first detection of GW \citep[e.g.,][]{2016ApJ...818L..22A, 2016Natur.534..512B, 2016MNRAS.461.3877D, 2017MNRAS.472.2422M, 2017NewA...51..122L, 2017MNRAS.471L.105S, 2017MNRAS.470.4739S, 2017NatCo...814906S, 2017ApJ...846...82Z, 2018MNRAS.481.4775D, 2018MNRAS.473.1186E, 2018ApJ...860....5G, 2018MNRAS.479.4391M, 2018PhRvD..97j3014S, 2018PhRvL.120o1101R, 2018ApJ...866L...5R, 2018PhRvD..98l3005R}. Below, we summarize those mechanisms for both sBBHs and BNSs.

For sBBHs, there are mainly four mechanisms, including (1) evolution of massive binary stars (hereafter denoted as EMBS channel; e.g., \citealt[][]{1973NInfo..27...70T, 1998ApJ...506..780B, 2002ApJ...572..407B, 2007ApJ...662..504B, 2008ApJS..174..223B, 2012ApJ...759...52D, 2013ApJ...779...72D, 2015ApJ...806..263D, 2015ApJ...814...58D, 2016Natur.534..512B, 2016MNRAS.461.3877D, 2016MNRAS.462.3302E, 2016PhRvD..93h4029R, 2016ApJ...824L...8R, 2016ApJ...832L...2R, 2017MNRAS.472.2422M, 2017NatCo...814906S, 2018MNRAS.473.1186E, 2018MNRAS.474.2959G, 2018MNRAS.480.2011G}); (2) dynamical interactions of compact (binary) stars in dense star clusters or galactic nuclei \citep[][]{1993Natur.364..423S, 2000ApJ...528L..17P, 2006ApJ...637..937O, 2016PhRvD..93h4029R, 2016ApJ...824L...8R, 2016ApJ...832L...2R, 2018MNRAS.481.4775D, 2018PhRvL.120o1101R, 2018ApJ...866L...5R, 2018PhRvD..98l3005R, 2018MNRAS.481.5445S} or the Lidov-Kozai mechanism for heirarchical triple systems \citep[e.g.,][]{2012ApJ...757...27A, 2016ApJ...828...77V, 2017ApJ...836...39S, 2018PhRvD..97j3014S} (hereafter dynamical channel); (3) AGN/MBH assisted formation mechanism (hereafter AGN/MBH-assisted channel; e.g., \citealt[][]{2017ApJ...835..165B, 2017MNRAS.464..946S, 2018ApJ...859L..25Y}); and (4) dynamical interactions of primordial black holes (hereafter PBH channel; e.g., \citealt{2017PhRvD..96l3523A, 2018ApJ...854...41K, 2018ApJ...864...61C}). For BNSs, they may be mostly formed from evolution of massive binary stars (the EMBS channel) and the contribution from the dynamical interactions in star clusters or other mechanisms may be negligible \citep[e.g.,][]{2018A&A...615A..91B, 2008MNRAS.386..553I, 2019arXiv191010740Y}.

Different mechanisms may result in different properties and merger rate of sBBHs. For example, sBBHs formed via the EMBS channel normally have small eccentricities \citep[][]{2002ApJ...572..407B, 2016MNRAS.461.3877D}, and the spins of their two components may be aligned and the resulting effective spin can be large \citep[e.g.,][but \citealt{2017arXiv170607053B}]{2016Natur.534..512B, 2016ApJ...818L..22A}. Those sBBHs formed via the dynamical channel are probably systematically heavier than those via the EMBS channel \citep[e.g.,][]{2018PhRvL.120o1101R, 2018PhRvD..98l3005R}, and may have high eccentricities \citep[][]{2016ApJ...818L..22A, 2018PhRvL.120o1101R, 2018PhRvD..98l3005R}, and the spins of their two components may be randomly oriented with respect to each other \citep[e.g.,][]{2016ApJ...818L..22A, 2016PhRvD..93h4029R}. The AGN/MBH-assisted channel may also result in heavy sBBHs \citep[e.g.,][]{2018ApJ...859L..25Y}. Different properties of those sBBHs at their formation time may lead to different properties of those sBBHs systems right before their final mergers. This can be reflected in distributions of some properties of sBBHs detected by LIGO and Virgo, e.g., effective spin (see \citealt{2016ApJ...818L..22A, 2017Natur.548..426F}) and eccentricity (see \citealt{2016PhRvD..93h4029R, 2018PhRvL.120o1101R}), and may also have some effects on the strength and shape of the combined GWB spectrum (as a function of frequency; \citealt[][]{2018PhRvL.120i1101A}). 

The merger rates of sBBHs and BNSs and their cosmic evolution have also been extensively studied in the past two decades and become a hot topic recently. Theoretical estimates for the local merger rate of sBBHs range from $\sim 0.1$ to $\sim 4000\gpcyr$ via the EMBS channel \citep[e.g.,][]{2016Natur.534..512B, 2017MNRAS.472.2422M, 2018MNRAS.480.2011G, 2018MNRAS.479.4391M, 2019MNRAS.486.2494G}, from $0.6$ to $20\gpcyr$ via the dynamical channel \citep[e.g.,][]{2016PhRvD..93h4029R, 2018MNRAS.480.5645H, 2018ApJ...866L...5R}, and are about $3-4\gpcyr$ via the AGN/MBH-assisted channel \citep[][]{2017MNRAS.464..946S,2019ApJ...876..122Y}, and the estimates for the merger rate of BNSs range from $5$ to $1000\gpcyr$ from the EMBS channel \citep[][]{2018MNRAS.479.4391M, 2019MNRAS.486.2494G}. Observational constraints on the local merger rates of sBBHs and BNSs have been recently obtained according to the GW detection by LIGO and Virgo, i.e., $56^{+45.0}_{-46.3}\gpcyr$ for sBBHs and $920^{+2920}_{-810}\gpcyr$ for BNSs \citep{2019PhRvX...9c1040A} at the $90\%$ confidence level, with uncertainties of an order of a magnitude. The large uncertainties in both the observational constraints and theoretical estimates on the local merger rate of sBBHs, at least, hinder robust conclusions on which mechanism dominates the origin of detected sBBHs. 

GWs radiated from numerous inspiralling-merging-ringdown sBBHs and BNSs combine together and form a stochastic GWB in LIGO/Virgo and LISA bands. Most previous studies on the stochastic GWB from compact binaries assumed  circular binaries \citep[e.g.,][]{2016MNRAS.461.3877D, 2016PhRvL.116m1102A, 2018PhRvL.120i1101A, 2018arXiv180910360C}. Part of the reason is that compact binary systems from the EMBS  channel, the dominant one for the formation of sBBHs and BNSs, though arguable, are believed to be effectively circularized well before they enter LIGO/Virgo and LISA bands.  However, the heavy sBBHs, such as GW\,150914 (with a total mass of $63M_\odot$) and GW\,170819 ($80M_\odot$), detected by LIGO/Virgo appear to be easier explained by the dynamic channel rather than by the EMBS ones \citep[e.g.,][]{2016AAS...22831509R, 2016ApJ...824L...8R}. Numerical calculations also suggest that the dynamical channel may contribute a non-negligible fraction to the BBH mergers, especially at the high-mass end \citep[e.g.,][]{2016PhRvD..93h4029R}. The AGN/MBH-assisted channel may also contribute to the formation sBBHs with extremely high eccentricities \citep[e.g.,][]{2019ApJ...877...87Z}. In these cases, a significant fraction of sBBHs may have significant eccentricities when they enter the LISA band, though they will be well circularized due to GW radiation before their final mergers.

In this paper, we estimate the strength and shape of the stochastic GWB emitted from the inspiral, merger, and ringdown of sBBHs and BNSs by considering different properties of those systems resulting from different formation channels. We focus at low frequencies from $10^{-4}-1$\,Hz that will be detected by LISA and Taiji/TianQin in future. We investigate whether different formation channels can be distinguished by using future observations on the stochastic low frequency GWB and illustrate the GWB signal by composing the GWs from more than millions of mock sBBHs and BNSs at different merging stages distributed over the cosmic time.

This paper is organized as follows. In Section~\ref{sec:Ogw}, we describe the frame work to obtain the energy density spectrum of the stochastic GWB combined from the GW emission from a large number of individual compact (eccentric) binaries over the cosmic time. We introduce simple models for the formation of sBBHs and BNSs in Section~\ref{sec:mr} by considering different formation channels for both sBBHs and BNSs. We present our results on the GWB obtained from those models in Section~\ref{sec:gwb}. We also show the GWB signals obtained by composing more than millions of cosmic sBBHs and BNSs in different merger stages in Section~\ref{sec:GWsignal}. Discussions are arranged in Section~\ref{sec:discussion}. Conclusions are given in Section~\ref{sec:conc}.

Through out this paper, we adopt the standard $\Lambda$CDM cosmology model with $H_0 = 67.9 {\rm km s^{-1} Mpc^{-1}}$, $\Omega_{\rm m} = 0.306$, $\Omega_{\rm k} = 0$, and $\Omega_{\Lambda} = 0.694$ \citep[obtained from][]{2016A&A...594A..13P}.
 
\section{GWB From Compact Binaries}
\label{sec:Ogw}

\subsection{GW radiation and orbital decay of eccentric binaries}
\label{sec:basic}

GW radiation causes the orbital decay of a compact binary and may lead to the final merger of the binary. This evolution process can be divided into three stages, i.e., inspiral, merger, and ringdown \citep{1998PhRvD..57.4535F}. In the inspiral stage, the period averaged evolution of the binary semimajor axis ($a$) and eccentricity ($e$) are given by \citep{1964PhRv..136.1224P}
\be 
\left \langle \frac{da}{dt} \right \rangle = - \frac{64}{5} 
\frac{G^3m_1m_2(m_1+m_2)}{c^5a^3}\frac{\left(1+\frac{73}{24}e^2
	+\frac{37}{96}e^4\right)}{(1-e^2)^{7/2}},
\label{eq:dadt}
\ee
and
\be 
\left\langle \frac{de}{dt} \right\rangle = - \frac{304}{15} \frac{G^3m_1m_2(m_1+m_2)}{c^5a^4}\frac{e\left(1+\frac{121}{304}e^2\right)}{(1-e^2)^{5/2}}.
\label{eq:dedt}
\ee
Here $G$ is the gravitational constant, $c$ is the speed of light, $m_1$ and $m_2$ represent the masses of the primary and secondary components, respectively. Almost all compact binary mergers (sBBHs and BNSs) are circularized (or at least close to circular orbits) well before they enter into the merger stage because of GW radiation. 

The orbital evolution due to GW radiation leads to a change of the orbital frequency $f_{\rm p}$ [$=G^{1/2}(m_1+m_2)^{1/2}a^{-3/2} /2\pi$] (a fundamental frequency of the system), which is described by the relationship  \citep[e.g.,][]{2007PThPh.117..241E}
\begin{equation}
\frac{f_{\rm p}}{f_{\rm p,0}}=\left[\frac{1-e_0^2}{1-e^2}\left(\frac{e}{e_0}\right)^{12/19}
\left(\frac{1+\frac{121}{304}e^2}{1+\frac{121}{304}e_0^2}\right)^{870/2299} \right]^{-3/2},
\label{eq:f2e}
\end{equation}
where $e_0$ and $f_{\rm p,0}$ are the initial eccentricity and orbital frequency of the binary, $f_{\rm p}$ is the orbital frequency when the orbital eccentricity evolves to $e$. According to the above Equation, for any given initial condition $(e_0, f_{{\rm p},0})$, the eccentricity $e(f_{\rm p}; e_0, f_{{\rm p},0})$ can be solved. 

Eccentric compact binaries emit GWs at the orbital frequency and all of its high order harmonics \citep{1964PhRv..136.1224P}. The GW energy density emitted from a distant eccentric binary during the inspiral stage at a given frequency $f_{\rm r}$ in the source's rest frame is given by \citep[e.g.,][]{PM63, 2015PhRvD..92f3010H, 2007PThPh.117..241E, 2017MNRAS.470.1738C}:
\begin{eqnarray}
\frac{d E_{\GW}(f_{\rm r})}{d f_{\rm r}} 
& = & \frac{(\pi G)^{2/3}}{3} \mc^{5/3} f_{\rm r}^{-1/3} 
\displaystyle\sum_{n=1}^{\infty} \left(\frac{2}{n}\right)^{2/3} \frac{g(n,e)}{F(e)}, \nonumber \\
& = & \frac{(\pi G)^{2/3}}{3} \mc^{5/3} f_{\rm r}^{-1/3} \Phi,
\label{eq:gw-ecc}
\end{eqnarray}
where 
\begin{eqnarray}
g(n,e)& = & \frac{n^4}{32} \left\{[J_{n-2}(ne)-2eJ_{n-1}(ne)+
\frac{2}{n}J_n(ne)\right. \nonumber \\
& &  + 2eJ_{n+1}(ne)-J_{n+2}(ne)]^2+(1-e^2) [J_{n-2}(ne)  \nonumber \\
& & \left. -2J_n(ne)+J_{n+2}(ne) ]^2+\frac{4}{3n^2}[J_n(ne)]^2 \right\} , \\
F(e)& = & \frac{1+(73/24)e^2+(37/96)e^2}{(1-e^2)^{7/2}},  \\
\Phi & = &\displaystyle\sum_{n=1}^{\infty} \left(\frac{2}{n}\right)^{2/3} \frac{g(n,e)}{F(e)} 
\label{eq:gne}
\end{eqnarray}
Here $J_n$ are Bessel functions, $\mc = (m_1m_2)^{3/5}/(m_1+m_2)^{1/5}$ is the chirp mass of the binary and $f_n= nf_{\rm p}$ is the GW frequency of the $n$-th harmonic in the source's rest frame, $e=e(f_{\rm r}; e_0,f_{n,0})$ is given by Equation~(\ref{eq:f2e}). According to Equation~(\ref{eq:gne}), the higher $e$ is, the higher harmonics contribute significant to the GW power.  

Equation~(\ref{eq:gw-ecc}) is reduced to the one for circular case ($e=0$) as $F(0)=1$, $g(2,0)=1$, and $g(n,0)=0$ for $n\neq 2$. In this case, GW is radiated at a single frequency $f_2=2f_{\rm p}$, twice of the orbital frequency $f_{\rm p}$, at any given time during the inspiral stage (as also described below in Eq.~\ref{eq:gw-circ}).

The GW energy spectrum for inspiral-merger-ringdown stages for a binary on a circular orbit can be described as \citep{2008PhRvD..77j4017A,2011ApJ...739...86Z}: 
\be 
\frac{dE_{\rm GW}}{df_{\rm r}}=\frac{(\pi G)^{2/3}\mc^{5/3}}{3}
 \begin{cases}
    f_{\rm r}^{-1/3},&  \textrm{if}\ f_{\rm r}\leq f_{\rm mrg},\\
    \omega_1 f_{\rm r}^{2/3},  & \textrm{if}\ f_{\rm mrg} < f_{\rm r} < f_{\rm rd},\\
    \frac{\omega_2 f_{\rm r}^2}{[1+4 (f_{\rm r}-f_{\rm rd})^2/\sigma^2]^2}, & 
    \textrm{if}\ f_{\rm rd} \leq f_{\rm r}\leq f_{\rm cut}.
\end{cases}
\label{eq:gw-circ}
\ee
Here $f_{\rm mrg}$ and $f_{\rm rd}$ represent GW frequency at the beginning of merger, and ring down stage, respectively, $f_{\rm cut}$ is the cutoff frequency of the template, and $\sigma$ is the width of a Lorentzian function defined in the third line at the r.h.s of the above equation, the coefficients $\omega_1 = f_{\rm mrg}^{-1}$ and $\omega_2= f_{\rm mrg}^{-1} f_{\rm rd}^{-4/3}$ are adopted to make the function $d E_{\GW}/d f$ continuous at frequencies $f_{\rm mrg}$ and $f_{\rm rd}$. Parameters $f_{\rm mrg}$, $f_{\rm rd}$, $f_{\rm cut}$, and $\sigma$ can be approximated as quadratic polynomials in terms of $\mc$ and $\eta$ ($=m_1m_2/(m_1+m_2)^2$, the symmetric mass ratio) of the hybrid waveforms as given in \citealt[][see their Eq.\,(4.18) and table I]{2008PhRvD..77j4017A}.
Since almost all eccentric compact binary mergers (sBBHs and BNSs) are circularized well before they enter into the merger stage, we adopt Equation~(\ref{eq:gw-circ}) to calculate the GW energy density for the merger and ring down stages for binaries even with high initial eccentricities.

The energy density per logarithmic frequency of the stochastic GWB from numerous inspiralling (eccentric) compact binaries can be written as  \citep[c.f.,][]{Phinney01}
\begin{eqnarray}
& & \Omega_{\rm{GW}}(f_{\rm o})  =  \frac{1}{\rho_{\rm c}}\frac{d\rho_{\GW}(f_{\rm o})}{d\ln f_{\rm o}} \nonumber \\
&  & =
\frac{8(\pi G)^\frac{5}{3}}{9c^2H_0^3}f_{\rm o}^\frac{2}{3}
\iiint d\mc de_0 dz \mc^{\frac{5}{3}}\frac{\Rmrg(\mc,e_0,z)}{(1+z)^\frac{1}{3}E_{\rm V}(z)} \Phi, \nonumber \\
\label{eq:gwdendf}
\end{eqnarray}
according to Equation~(\ref{eq:gw-ecc}). Here $\Rmrg(\mc,e_0,z)$ is the cosmic merger rate density at redshift $z$ of compact binaries with chirp mass $\mc$ (determined by the total mass and mass ratio),
$f_{\rm o}$ [$=f_{\rm r}/(1+z)$] is the GW frequency at the observer's rest frame, $E_{\rm V}(z)=(1+z) \sqrt{\Omega_{\rm m}(1+z)^3+\Omega_\Lambda}$, $\rho_{\rm c}= 3c^2H_0^2/8\pi G$ is the critical comoving density of the universe and $H_0$ is the Hubble constant. 

To obtain the GWB, it is necessary to include high harmonic GW radiation from eccentric binaries. In general it is sufficient to sum $n$ up to a value $n_{\rm max}$ if the contribution of $n> n_{\rm max}$ harmonics is negligible to the total GW energy spectrum. For a given eccentricity, we adopt $n_{\rm max} \sim 10n_{\rm peak}$ with $n_{\rm peak} \approx \frac{2(1+e)^{1.1954}}{(1-e^2)^{3/2}}$ the harmonics that contribute the most to the GW density spectrum \citep{2003ApJ...598..419W,2010PhRvD..82j7501B}. 

Equation~(\ref{eq:gwdendf}) can be reduced to a simpler one if all binaries are circularized. Assuming circular binaries, the GWB energy density can be obtained by
\begin{eqnarray} 
\Omega_{\rm{GW}}(f_{\rm o}) =  \frac{8(\pi G)^\frac{5}{3}}{9c^2H_0^3}
\iint d\mc dz \mc^{\frac{5}{3}} \frac{\Rmrg(\mc, 0, z)}{E_{\rm V}(z)} \times \nonumber  \\
\begin{cases}
   f_{\rm o}^{2/3} /(1+z)^{1/3}, &  \textrm{if}\ f_{\rm r}\leq f_{\rm mrg},  \\
  \omega_1 f_{\rm o}^{5/3}(1+z)^{2/3},  & \textrm{if}\ f_{\rm mrg} < f_{\rm r} < f_{\rm rd},  \\
    \frac{\omega_2 f_{\rm o}^3(1+z)^2}{[1+4 (f_{\rm o}(1+z)-f_{\rm rd})^2/\sigma^2]^2}, & 
    \textrm{if}\ f_{\rm rd} \leq f_{\rm r}\leq f_{\rm cut}. 
\end{cases}
\label{eq:gwdendf1}
\end{eqnarray}

\subsection{Merger Rates}
\label{sec:mr}

The GWB depends on the cosmic merger rate densities of GW sources, either sBBHs or BNSs. Below we describe our simple estimates on the cosmic merger rate densities of sBBHs and BNSs resulting from different formation mechanisms/channels. The total merger rate density $\Rmrg(m_1,q,e_0,z) \simeq \Rmrgb(m_1,q,e_0,z)+\Rmrgd(m_1,q,e_0,z)+\Rmrgns(m_1,q,e_0,z)$, where $\Rmrgb(m_1,q,e_0,z)$ and $\Rmrgd(m_1,q,e_0,z)$ are the merger rate density of sBBHs at redshift $z$ from the EMBS channel and the dynamical channel, respectively, $m_1$, $q$, and $e_0$ are the primary mass, mass ratio, and eccentricity $e_0$ at a given frequency $f_0$ of the binary, and $\Rmrgns$ is the merger rate density for BNSs from the EMBS channel. The merger rate can also be converted to $\Rmrg(\mc,e_0,z)$ with $\mc= q^{3/5} m_1/(1+q)^{1/5} $. To simplify the problem, we assume that the eccentricity distribution can be separated at the given frequency $f_0$, i.e., independent of $\mc$ and $z$. Therefore the eccentricity distribution can be first ignored when we estimate $\Rmrgb(m_1,q,e_0,z)$, $\Rmrgd(m_1,q,e_0,z)$, and $\Rmrgns(m_1,q,e_0,z)$ below, and then taken into account when estimating the GWB.

\subsubsection{Cosmic Merger Rate Density of sBBHs and sBBH properties}
\label{sec:mr_bbh}

sBBHs can be formed via four different channels as summarized in Section~\ref{sec:intro}. In this work, we mainly consider the first two channels, but ignore those sBBHs formed via either the AGN/MBH-assisted channel or the PBH channel. We neglect sBBHs from the AGN/MBH-assisted channel since this channel may only lead to a sBBH merger rate substantially smaller than that from the EMBS and dynamical channels \citep[e.g.,][]{2017MNRAS.464..946S, 2019ApJ...877...87Z}.\footnote{Note here that sBBHs formed via the AGN/MBH-assisted channel may also have large eccentricities \citep[e.g.,][]{2019ApJ...877...87Z} and thus have an effect similar to those formed via the dynamical channel on the GWB. In principle, its effect can be absorbed into that from the dynamical channel by adjusting the contribution fraction. } The merger rate of sBBHs from the PBH channel is highly uncertain, although it is argued that the contribution from it to the GWB may be also large comparing with that from the astrophysical channels \citep[e.g.,][]{2018arXiv180910360C}. In the present paper, we also ignore the contribution from the primordial channel.

\begin{itemize}
\item {\bf EMBS channel:} The cosmic merger rate density for sBBHs formed via the EMBS channel may be simply derived by the convolution of the birth rate density of sBBHs with the distribution function of the time delay ($t_{\rm d}$) of the merger time from the formation time as \citep[see][]{2016MNRAS.461.3877D, 2018MNRAS.474.4997C}:
\be
\Rmrgb(m_1, q, z) = \int dt_{\rm{d}} f_{\rm eff}R_{\rm birth}(m_1, z')P_t(t_{\rm d})P_q(q) ,
\label{eq:Rmrg}
\ee
and
\be
R_{\rm birth}(m_1,z') = \iint dm_\star dZ \dot{\psi}(Z;
z')\phi(m_\star) \delta (m_\star - g^{-1}(m_1,Z)).
\label{eq:Rbir}
\ee

Here $\dot{\psi}(Z; z')$ is the cosmic star formation rate density (SFR) with metallicity $Z$ in the range $Z \rightarrow Z+dZ$ at redshift $z'$, $\phi(m_\star)$ is the initial mass function (IMF) and $m_1 = g(m_\star, Z)$ is the relationship between the BH remnant mass $m_1$ and its progenitor stellar mass $m_\star$, for which we adopt the results by \citet{2015MNRAS.451.4086S}, the probability distribution of $t_{\rm d}$ ($=t(z)-t(z')$ with $t(z) =\int_z^{\infty} |dt/dz'| dz'$) is assumed to $P(t_{\rm d})\propto t_{\rm d}^{-1}$, with the minimum value as $50$\,Myr and maximum value as the Hubble time  \citep{2016Natur.534..512B, 2016MNRAS.461.3877D}, $P_{q}(q) \propto q$, that ranges from $0.5$ to $1$, is the distribution of mass ratio $q$ ($=m_2/m_1$). We assume that $\dot{\psi}(Z;z)$ can be separate to two independent functions, one is the total SFR at redshift $z$ and the other is metallicity distribution of those stars at that redshift. We adopt the total SFR obtained from observations by \citet{2014ARA&A..52..415M} as
\be
{\rm SFR}(z) = 0.015\frac{(1+z)^{2.7}}{1+[(1+z)/2.9]^{5.6}} M_\odot{\rm
Mpc^{-3}\,yr^{-1}},
\label{eq:sfr}
\ee
and mean metallicity distribution \citep{2016Natur.534..512B} 
\be
\log\left[Z_{\rm mean}(z)\right]=0.5+\log\left( \frac{y(1-R)}{\rho_{\rm
b}}\int_{z}^{20}\frac{97.8 \times10^{10}{\rm SFR}(z^\prime)}{H_0
E_{\rm V}(z^\prime)(1+z^\prime)}dz^\prime \right).
\label{eq:Metallicity}
\ee
Here $R=0.27$, $y=0.019$, $\rho_{\rm b}=2.55\times10^{11}\, \Omega_{\rm b}\,h_0^2\,M_\odot{\rm Mpc^{-3}}$, $h_0$ is the Hubble constant in unit of $100{\rm km\,s^{-1} Mpc^{-1}}$.
More detailed descriptions about the method to estimate the sBBH merger rate density can be found in \citet{2018MNRAS.474.4997C}. 

sBBHs formed via the EMBS channel may also be initially eccentric due to the natal kick received at the formation time of the second BH, which may lead to an imprint in the shape of the GWB spectrum at the LISA band.
During the binary evolution, the eccentricity introduced by natal kick and mass loss can be estimated by the descriptions in \citet{2002MNRAS.329..897H}. Observations of single Galactic pulsars suggest that the natal kick is about $265\,{\rm km\,s^{-1}}$ \citep{2005MNRAS.360..974H}, while the natal kicks should be lower for close binaries. Assuming that the natal kick is $200\,{\rm km\,s^{-1}}$ and the mass loss at black hole formation is $10\%$, the induced eccentricity can be $\sim0.3$, which is consistent with the result in \citet{2011A&A...527A..70K}. Apparently, the eccentricities excited by the natal kick cannot be very high \citep[e.g.,][]{2011A&A...527A..70K,2016Natur.534..512B} for any reasonably assumed natal kicks. In the present paper, we consider two cases for the eccentricity of the sBBHs formed via the EMBS channel, i.e., (1) all sBBHs are already well circularized and have eccentricities close to $0$ when they enter the LISA band from $f_{\rm r}=10^{-4}$\,Hz, and (2) the eccentricity probability distribution of those sBBHs is Gaussian with mean eccentricity $0.3$ and standard deviation $0.1$ when those sBBHs enter the LISA band (i.e., $e \sim \rm N$ ($0.3,0.1^2$) at $10^{-4}$\,Hz). 

\item {\bf Dynamical channel:} The formation of sBBHs via dynamical interactions in dense (globular) clusters have been investigated intensively and its cosmic merger rate density has been estimated in a number of recent works. In this paper, we adopt the merger rate density obtained in \citet[][see their Eq.\,(1) and Appendix]{2018ApJ...866L...5R} by using both dynamical simulations on the formation of sBBHs and simple descriptions on the formation and evolution of globular clusters, which is given by
\begin{eqnarray}
 \Rmrgd(t) = &\iiint \left. \frac{ \dot{M}_{\rm GC}} {d\log_{10}M_{\rm    
	 Halo}} \right|_{z(\tau)} \frac{1}{\left< M_{\rm GC}\right>} 
	 P(M_{\rm GC}) \nonumber\\
	 & \times R(r_{\rm v},M_{\rm GC},\tau-t) dM_{\rm{Halo}}dM_{\rm GC} d\tau~.
 \label{eqn:master}
\end{eqnarray}
Here $\frac{\dot{M}_{\rm GC}} {d\log_{10}M_{\rm Halo}}$ is the comoving SFR in globular clusters per galaxies of a given halo mass $M_{\rm Halo}$ at given redshift $z(\tau)$ (or a given formation time $\tau$), $P(M_{\rm GC})$ is the cluster initial mass function, $\left< M_{\rm GC}\right>$ is the mean initial mass of a globular cluster and $R(r_{v},M_{\rm GC},t)$ is the merger rate of sBBHs in a globular cluster with initial virial radius $r_{v}$ and mass $M_{\rm GC}$ at time $t$. The specific fit for $\frac{\dot{M}_{\rm GC}} {d\log_{10}M_{\rm Halo}}$ and $R(r_{v},M_{\rm GC},t)$ can be found in the Appendix of \citet{2018ApJ...866L...5R}. Here we adopt their standard model, which assumes $50\%$ of clusters form with $r_{\rm v} = 1 {\rm pc}$ and $50\%$ form with $r_{\rm v} = 2 {\rm pc}$. 

For sBBHs formed via the dynamical channel, we can get the distribution of their total mass and mass ratio ($q$), according to the simulation results in \citet{2018PhRvD..98l3005R, 
2016PhRvD..93h4029R}. The resulting mass ratio distribution ($P(q)$) is more or less similar to that from the EMBS channel. Therefore, we adopt the same distribution as that for the EMBS channel sBBHs, i.e., $P(q) \propto q$ for $q\in [0.5,1]$. We obtain the primary mass distribution $P(m_1)$ using the total mass and mass ratio distributions, then we get $\Rmrgd(m_1,q,z)$ for the dynamical channel. 

The sBBHs formed via the dynamical channel in dense globular cluster may have large eccentricities when they radiate GWs in the LISA band \citep{2018PhRvL.120o1101R, 2018MNRAS.481.4775D}. In general, the dynamically originated sBBHs experienced encounters with other objects for many times excited to high eccentric orbits \citep{1996MNRAS.282.1064H}, and thus they are expected to possess eccentricities much larger than those formed via the EMBS channel. According to the simulation results obtained in many recent works \citep{2016ApJ...830L..18B, 2018MNRAS.481.5445S, 2018PhRvD..98l3005R}, usually the eccentricity can be as high as $10^{-4}$ at $10$Hz. According to \citet{2016ApJ...830L..18B}, the eccentricities of dynamically originated sBBHs can be as large as $0.9$ or even larger at $10^{-4}$Hz, and more than half of them can have such eccentricities. In the present paper, we assume three different eccentricity distributions for sBBHs formed via the dynamical channel, first one is a Gaussian distribution at orbital frequencies of $10^{-4}$\,Hz with mean of $0.9$, second one is a Gaussian distribution at the orbital frequency of $10^{-3}$\,Hz with mean of $0.7$, and the last one is a uniform distribution between $0.5$ and $1$ at the orbital frequency of $10^{-3}$\,Hz. Since $e\leq 1$ and $\geq 0$, we cut the distribution at $e=1$ and $0$, and renormalize it accordingly. We note here that the sBBHs formed from AGN/MBH-assisted channel and those affected by Lidov-Kozai mechanism can have even larger eccentricities. Their contribution to the GWB may further strengthen the effects on GWB shape by dynamical originated sBBHs discussed below.
\end{itemize}

\begin{figure}
\centering
\includegraphics[scale=0.6]{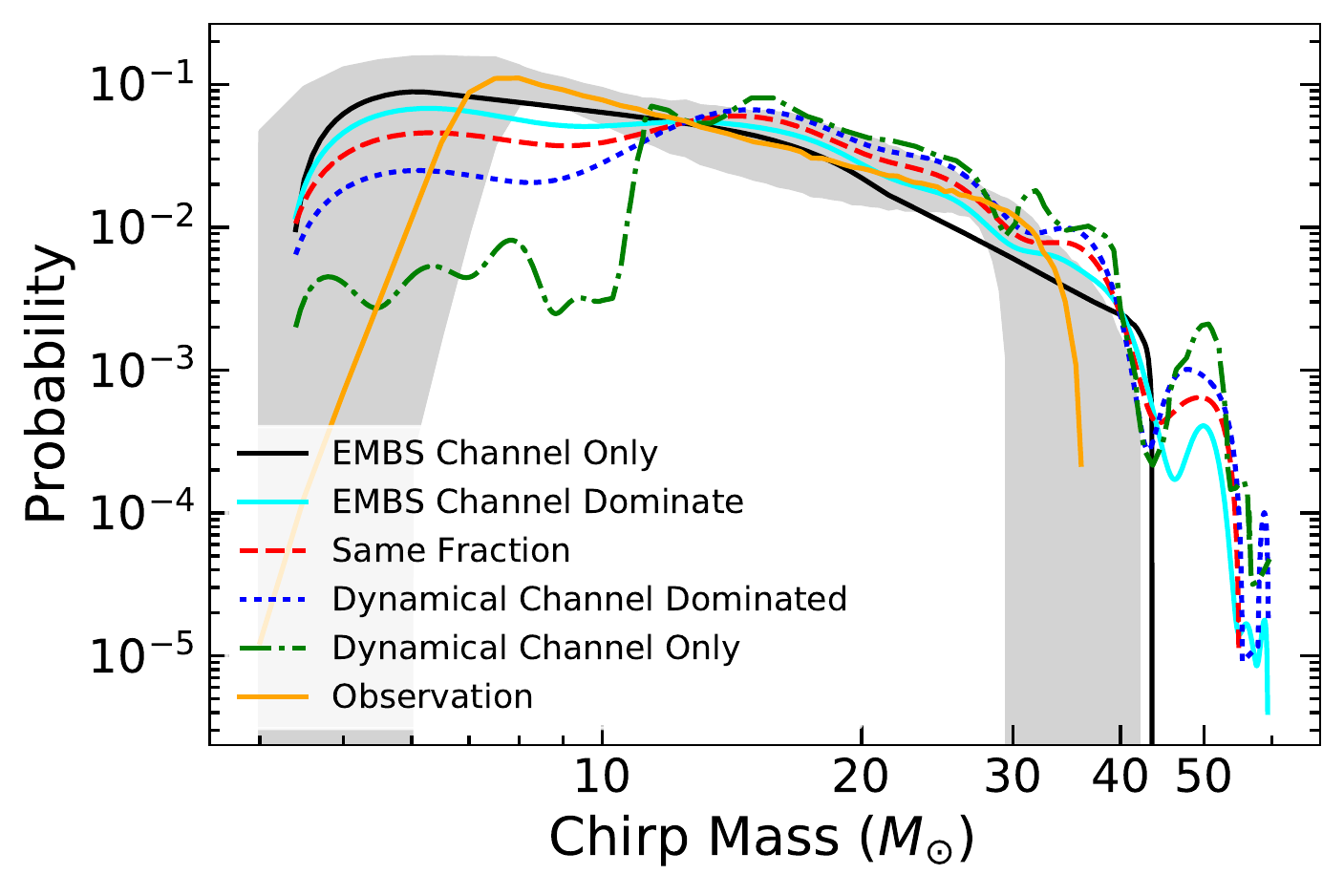}
\caption{Chirp mass distributions of sBBHs in different models at redshift $z = 0$. Black solid line represents the result from the model where EMBS channel produces all sBBHs (see the first line of Table~\ref{tab:model}). Cyan, red dash, blue dot, and green dotted-dash lines show the results from the model with contributions from the dynamical channel of $25\%$, $50\%$, $75\%$, and $100\%$, respectively. The shaded region represents the constraint from LIGO/VIRGO O1 and O2 observations \citep{2019ApJ...882L..24A}.
}
\label{fig:f1}
\end{figure}

\subsubsection{Cosmic merger rate density of BNSs and BNSs properties}
\label{sec:mr_BNS}

Similar to the description of merger rate density of sBBHs formed from the EMBS channel, we use an analytic description to calculate the cosmic merger rate density for BNSs. This description is similar to that for BBHs represented by Equations~\eqref{eq:Rmrg} and \eqref{eq:Rbir}, except that the mass range for $m_1$ of BNSs is from $1 M_{\odot}$ to $2 M_{\odot}$. We also adopt the relation between remnant mass and progenitor star mass from \citet{2015MNRAS.451.4086S}, the same as what we use for sBBHs.

These BNSs are eccentric at formation and we assume a simple Gaussian distribution, $P_{\rm BNS}(e) = N(0.7,0.1^2)$ at orbital frequency of $10^{-4}$ Hz, to describe \citep{2020ApJ...892L...9A}. The eccentricity distribution may have a large uncertainty as different formation models may result quite different distributions \citep{2018A&A...619A..77K}.

\subsection{Models}

Here we introduce twelve different models to estimate the GWB by considering the uncertainties in the estimates of sBBH merger rates for different formation channels and the eccentricity distribution of those sBBHs/BNSs, as listed in Table~\ref{tab:model}. The total sBBH and BNS merger rates at $z\sim 0$ for each model are calibrated to the current constraints obtained from the O1 and O2 observations of LIGO and VIRGO. These constraints on the local merger rates densities for sBBHs and BNSs are $56^{+45.0}_{-46.3} \gpcyr$ and $920^{+2920}_{-810} \gpcyr$, respectively \citep{2019PhRvX...9c1040A}.\footnote{Considering recent detection of GW190425, the latest constraint on the BNS local merger rate is consistent with the value adopted here \citep{GW190425}.} Descriptions of those sBBHs contributed from the EMBS channel and the dynamical channel in each model are itemized as follows. For all these models, the mergers from BNSs are fixed as described in Section~\ref{sec:mr_BNS}  and their contribution to the GWB is the same.

\begin{enumerate}
    \item{\bf R1:0e1\_d0e2\_...:}
    In this model, we assume all the sBBHs are originated from EMBS channel. The redshift evolution of their cosmic merger rate density is described in section \ref{sec:mr_bbh}. We assume these sources are in circular orbits, meaning their eccentricity ($e_1$) probability distribution is $P(e_1) =\delta(0)$.

    \item {\bf R3:1e1\_d0e2-4\_g9:}
    In this model, we adopt the local sBBH merger rate density from the dynamical channel as $14 \gpcyr$ \citep{2018ApJ...866L...5R}, i.e., accounting for $25\%$ of all the sBBHs. The rest $75\%$ local sBBH mergers are from the EMBS channel. The ratio of the local sBBH merger rate contributed from the EMBS channel to that from the dynamical channel is $R=3:1$. The redshift evolution of the EMBS sources and the dynamical sources are the same as those given by Equation~(\ref{eq:Rmrg}) and (\ref{eqn:master}). We assume the EMBS sources are 
    all in circular orbits, i.e., their eccentricity ($e_1$) probability distribution is $P(e_1) =\delta(0)$. For dynamical origin sources their eccentricity distribution $P(e_2)$ is assumed to be a Gaussian distribution with a mean of $0.9$ and a standard deviation of $0.1$ at $10^{-4}$,Hz, denoted as $P(e_2) = \left. N(0.9,0.1^{2})\right|_{10^{-4}{\rm Hz}}$. This model is frequently taken as the reference model in the text below.

    \item{\bf R3:1e1\_g3e2-4\_g9:}
    The settings of this model are the same as that of the model {\bf R3:1e1\_d0e2-4\_g9} except that the eccentricity distribution of the EMBS sources is assumed to follow a Gaussian distribution with a mean of $0.3$ and a standard deviation of $0.1$ at orbital frequency of $10^{-4}$\,Hz, i.e., $P(e_1)= \left. N(0.3,0.1^{2})\right|_{10^{-4}{\rm Hz}}$.
    
    \item{\bf R3:1e1\_d0e2-3\_g7:}
    The settings of this model are the same as that of the model {\bf R3:1e1\_d0e2-4\_g9} except that the eccentricity distribution of the dynamical sources is assumed to follow a Gaussian distribution with a mean of $0.7$ and a standard deviation of $0.1$ at orbital frequency of $10^{-3}$\,Hz, i.e., $ P(e_2) = \left. N(0.7,0.1^{2})\right|_{10^{-3}{\rm Hz}}$.
    
    \item{\bf R3:1e1\_g3e2-3\_g7:}
    The settings of this model are the same as that of the model {\bf R3:1e1\_d0e2-3\_g7} except that the eccentricity distribution of the EMBS sources is assumed to follow a Gaussian distribution with a mean of $0.3$ and a standard deviation of $0.1$ at orbital frequency of $10^{-4}$\,Hz, i.e.,$ P(e_1) = \left. N(0.3,0.1^{2})\right|_{10^{-4}{\rm Hz}}$.
    
    \item{\bf R1:1e1\_d0e2-4\_g9:}
    The settings of this model are the same as that of the model {\bf R3:1e1\_d0e2-4\_g9} except that the ratio of local sBBH mergers contributed by the EMBS channel to that by the dynamical channel is set to 1:1.
    
    \item{\bf R1:1e1\_d0e2-3\_g7:}
    The settings of this model are the same as that of the model {\bf R1:1e1\_d0e2-4\_g9} except that the eccentricity distribution of the dynamical sources is assumed to follow a Gaussian distribution $P(e_2)=N(0.7,0.1^2)$ at orbital frequency of $10^{-3}$\,Hz.
    
    \item{\bf R1:3e1\_d0e2-4\_g9:}
    The settings of this model are the same as that of the model {\bf R1:1e1\_d0e2-4\_g9} except that the ratio of the local sBBH mergers contributed by the EMBS channel to that by the dynamical channel is set to $1:3$.
    
    \item{\bf R1:3e1\_d0e2-3\_g7:}
    The settings of this model are the same as that of the model {\bf R1:1e1\_d0e2-3\_g7} except that the ratio of the local sBBH mergers contributed by the EMBS channel to that by the dynamical channel is set to $1:3$.
    
    \item{\bf R0:1e1\_... e2-4\_g9:}
    In this model, sBBH mergers are assumed to be all formed from the dynamical channel with an eccentricities following a Gaussian distribution with a mean of $0.9$ and a standard deviation of $0.1$ at orbital frequency of $10^{-4}$\,Hz, i.e., $P(e_2)=\left. N(0.9,0.1^2) \right|_{10^{-4}{\rm Hz}}$.

    \item{\bf R0:1e1\_... e2-3\_g7:}
    In this model, sBBH mergers are assumed to be all formed from the dynamical channel with an eccentricities following a Gaussian distribution with a mean of $0.7$ and a standard deviation of $0.1$ at orbital frequency of $10^{-3}$\,Hz, i.e., $P(e_2)=\left. N(0.7,0.1^2) \right|_{10^{-3}{\rm Hz}}$. 
	\item{\bf R0:1e1\_...e2-3\_U(0.5,1):}
	In this model, sBBH mergers are assumed to be all formed from the dynamical channel with an eccentricities following a uniform distribution between $0.5$ and $1$ at orbital frequency of $10^{-3}$\,Hz, i.e.,$P(e_2)=\left. \rm U(0.5,1)\right|_{10^{-3}{\rm Hz}}$.
   
Note that we neglect the contribution from AGN/MBH-assisted channel in all the above models. This contribution and its consequent effect on the GWB shape may be absorbed into those models for the dynamical channel with extreme settings on the eccentricity distribution.

\end{enumerate}

Different models may result in sBBHs with different chirp mass distributions. As an example, Figure~\ref{fig:f1} shows the normalized chirp mass distribution functions at redshift $z=0$ from five different models. As seen from Figure~\ref{fig:f1}, the chirp masses of dynamical channel sources are mostly larger than $10 M_{\odot}$, while a significant fraction of the EMBS channel sources have chirp masses $\lesssim 10 M_\odot$. Here we only show the results at $\rm z = 0$, as for other redshifts the difference between the chirp mass distributions for these two channels remains more or less the same. The chirp masses of the dynamical origin sBBHs are relatively larger comparing with those from the EMBS channel. The chirp mass function from the model with $100\%$ sBBHs originated from the dynamical channel (green dotted-dash line) is top heavy with only a small fraction of sBBHs with $\mc\lesssim 10M_\odot$, while the model with $100\%$ sBBHs originated form the EMBS channel has a significant fraction of sBBHs with chirp mass $\mc \lesssim 10M_\odot$.   
The sBBH chirp mass distribution function resulting from the first model is quite consistent with the constraint obtained from the O1 and O2 LIGO/VIRGO observations (shaded region in Fig.~\ref{fig:f1}; see \citet{2019ApJ...882L..24A}). The chirp mass distribution function from the model with $100\%$ sBBHs from the dynamical channel seems inconsistent with the current observational constraint. However, in the present work, this model is still taken as an extreme case to demonstrate the effect of sBBH eccentricities on the shape of the GWB spectrum.  Note  that the  wiggle features in the chirp mass distribution curves shown in Figure~\ref{fig:f1} are due to that we adopt the simulation results from \citet{2018PhRvD..98l3005R} for dynamical originated sBBHs which is limited by the small number of sources with discrete masses.

\begin{table*}
\centering
\caption{
Parameter settings of sBBH models and the best fits to the resulting GWB at the band of LISA-like space GW detectors.
}
\label{tab:model}
\begin{tabular}{c|cc|cc|cc|c|cc} %
\hline		
\multirow{2}{*}{Model} & \multicolumn{2}{c}{sBBH Fraction}  & \multicolumn{2}{c}{sBBH $P(e)$ }  & $\Omega_{\rm GW}$(25Hz)  & $A$ &  $f_*$ & \multirow{2}{*}{$\beta$} \\ \cline{2-3}  \cline{4-5} \cline{6-8} 
		&  EMBS  & Dyna & EMBS  & Dyna & $10^{-9}$ &  $10^{-12}$ &  $10^{-3}$Hz & & \\ \hline
{\bf R1:0e1\_d0e2\_...} & $100\%$ & $0$  & $\delta(0)$ & $\cdots\cdots$  & $1.68$  & $1.00$  & $0.36$ & $0.88$ \\ \hline
{\bf R3:1e1\_d0e2-4\_g9} & $75\%$  & $25\%$ & $\delta(0)$& $\left. (0.9,0.1^2)\right|_{f_1}$ & $1.86$  & $1.36$  & $0.52$ & $0.89$\\ \hline
{\bf R3:1e1\_g3e2-4\_g9} & $75\%$ & $25\%$ & $\left. (0.3,0.1^2)\right|_{f_1}$ &  $\left. (0.9,0.1^2)\right|_{f_1}$ & $1.86$  & $1.28$ & $0.48$ & $0.86$\\ \hline
{\bf R3:1e1\_d0e2-3\_g7} & $75\%$ & $25\%$ & $\delta(0)$ & $\left. (0.7,0.1^2)\right|_{ f_2}$ & $1.86$ &  $2.22$  & $1.1$ & $0.83$\\ \hline
{\bf R3:1e1\_g3e2-3\_g7} & $75\%$ & 25\% & $\left. (0.3,0.1^2)\right|_{f_1}$ & $\left. (0.7,0.1^2)\right|_{ f_2}$ & $1.86$  &  $2.24$  & $1.1$ & $0.79$\\ \hline
{\bf R1:1e1\_d0e2-4\_g9} & $50\%$ & $50\%$ & $\delta(0)$ & $\left. (0.9,0.1^2)\right|_{ f_1}$ & $2.03$ &  $1.74$  & $0.68$ & $1.04$ \\ \hline
{\bf R1:1e1\_d0e2-3\_g7} & $50\%$ & $50\%$ & $\delta(0)$ & $\left. (0.7,0.1^2)\right|_{ f_2}$ & $2.03$ &  $3.49$ &  $1.9$ & $0.89$\\	\hline
{\bf R1:3e1\_d0e2-4\_g9} & $25\%$ & $75\%$ & $\delta(0)$ & $\left. (0.9,0.1^2)\right|_{ f_1}$ & $2.20$ &  $2.29$ &  $0.92$ & $1.10$ \\ \hline
{\bf R1:3e1\_d0e2-3\_g7} & $25\%$ & $75\%$ & $\delta(0)$ & $\left. (0.7,0.1^2)\right|_{ f_2}$ & $2.20$ &  $5.07$ &  $3.0$ & $0.95$\\ \hline
{\bf R0:1e1\_... e2-4\_g9} & $0$     & $100\%$ & $\cdots\cdots$ & $\left. (0.9,0.1^2)\right|_{ f_1}$ & $2.38$ &  $10.0$   & $3.2 $ & $1.20$\\ \hline
{\bf R0:1e1\_... e2-3\_g7} & $0$     & $100\%$ & $\cdots\cdots$ & $\left. (0.7,0.1^2)\right|_{ f_2}$ &  $2.38$ &  $10.0$  & $3.2$ & $1.35$\\ \hline
{\bf R0:1e1\_... e2-3\_U(0.5,1)} & $0$	& $100\%$  & $\cdots\cdots$ & $\left.\rm U(0.5,1)\right|_{ f_2}$ & $2.38$ & $10.0$ & $3.2$ & $1.25$ \\ \hline
\end{tabular}
\begin{flushleft}
\footnotesize{Note: first column denotes the model name, second and third columns show the local merger rate fractions from the EMBS and dynamical channels, respectively, fourth and fifth columns list the eccentricity distribution $P(e)$ of the EMBS and dynamical sources at an orbital frequency of either $f_1=10^{-4}$\,Hz or $f_2=10^{-3}$\,Hz (indicated by the subscript), respectively. In the fourth and fifth column, $(\bar{e}, \sigma^2)|_f$ represents a Gaussian distribution with a mean of $\bar{e}$ and a standard deviation of $\sigma$, $\delta$ is the Dirac function and $\rm U(0.5,1)$ means a uniform distribution between $0.5$ and $1$. Sixth column shows the GWB density at $25$\,Hz. Last three columns described the fitting results for the shape $\Omega_{\rm GW}$ by a double power-law, where $A$, $f_*$, and $\beta$ are the amplitude, bending frequency, and low-frequency power index, respectively (see Eq.~\ref{eq:fit} in Section~\ref{sec:gwb})}.
\end{flushleft}
\end{table*}

\subsection{GWB from Different Models}
\label{sec:gwb}

\begin{figure}
\begin{center}
\includegraphics[scale=0.56]{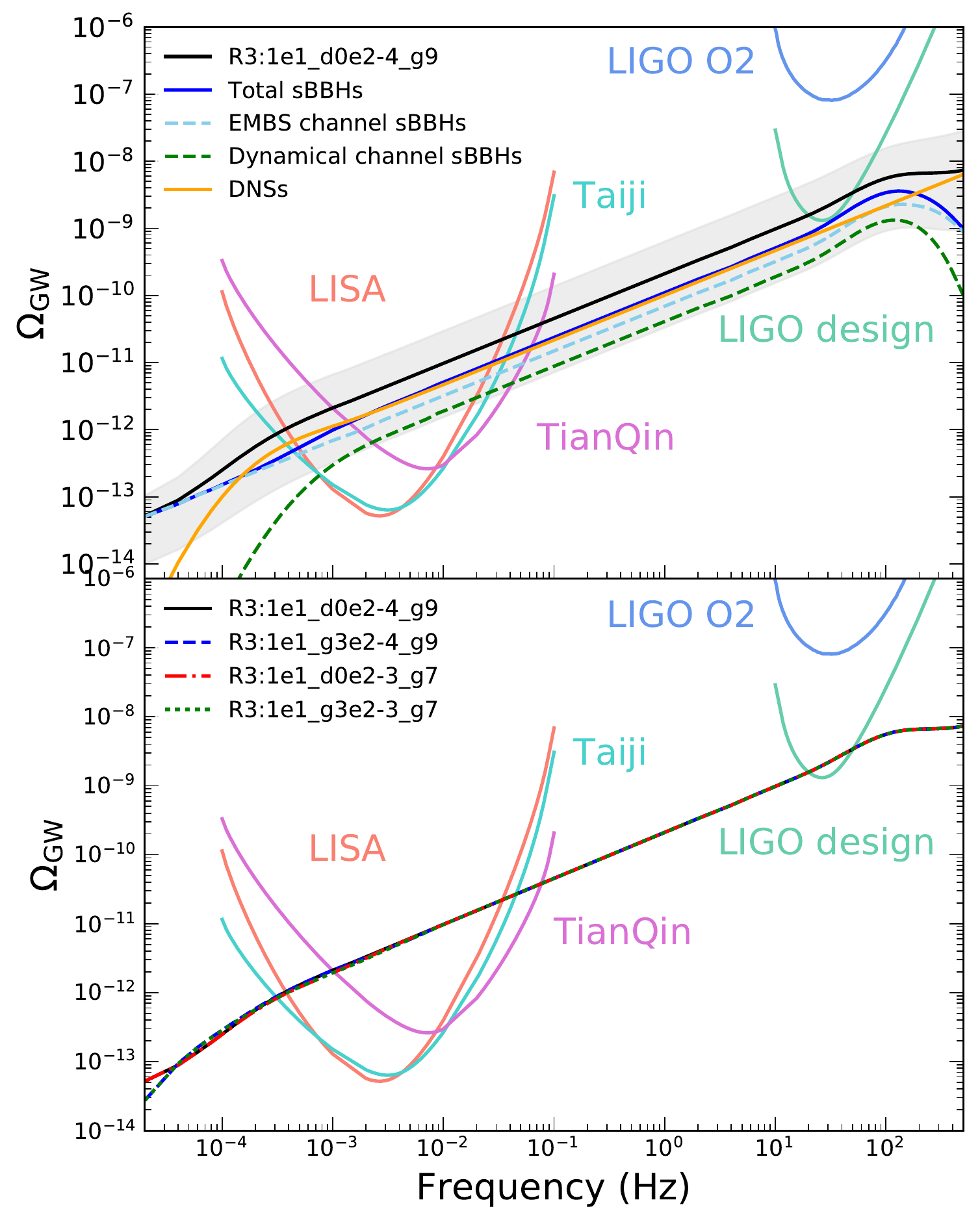}
\caption{
The energy density spectrum of stochastic GWB from stellar binary black holes (sBBHs) and binary neutron stars (BNSs) under different models. The top panel shows the result from {\bf R3:1e1\_d0e2-4\_g9} model. Black solid line represents the total $\Omega_{\rm GW}$, blue solid line represents the contribution from all the sBBHs and orange solid line represent the contribution from BNSs. Light blue dashed line shows $\Omega_{\rm GW}$ due to sBBHs from EMBS channel which accounts for $75\%$ of all the sBBHs. $\Omega_{\rm GW}$ from the rest $25\%$ dynamical sBBHs is shown by green dashed line. The bottom panel shows the total $\Omega_{\rm GW}$ from four different models with the same fractions of different sBBHs channels. The detailed information of these models is listed in Table~\ref{tab:model}. The differences between these models are the eccentricity distributions for different channels of sBBHs. The differences in the resulting $\Omega(f)$ from these four models are small. Red, cyan, violet, blue, and green curves show the sensitivity curves of LISA, Taiji, TianQin, LIGO O2, and LIGO design, respectively.
}
\label{fig:f2}
\end{center}
\end{figure}

\begin{figure}
\begin{center}
\includegraphics[scale=0.6]{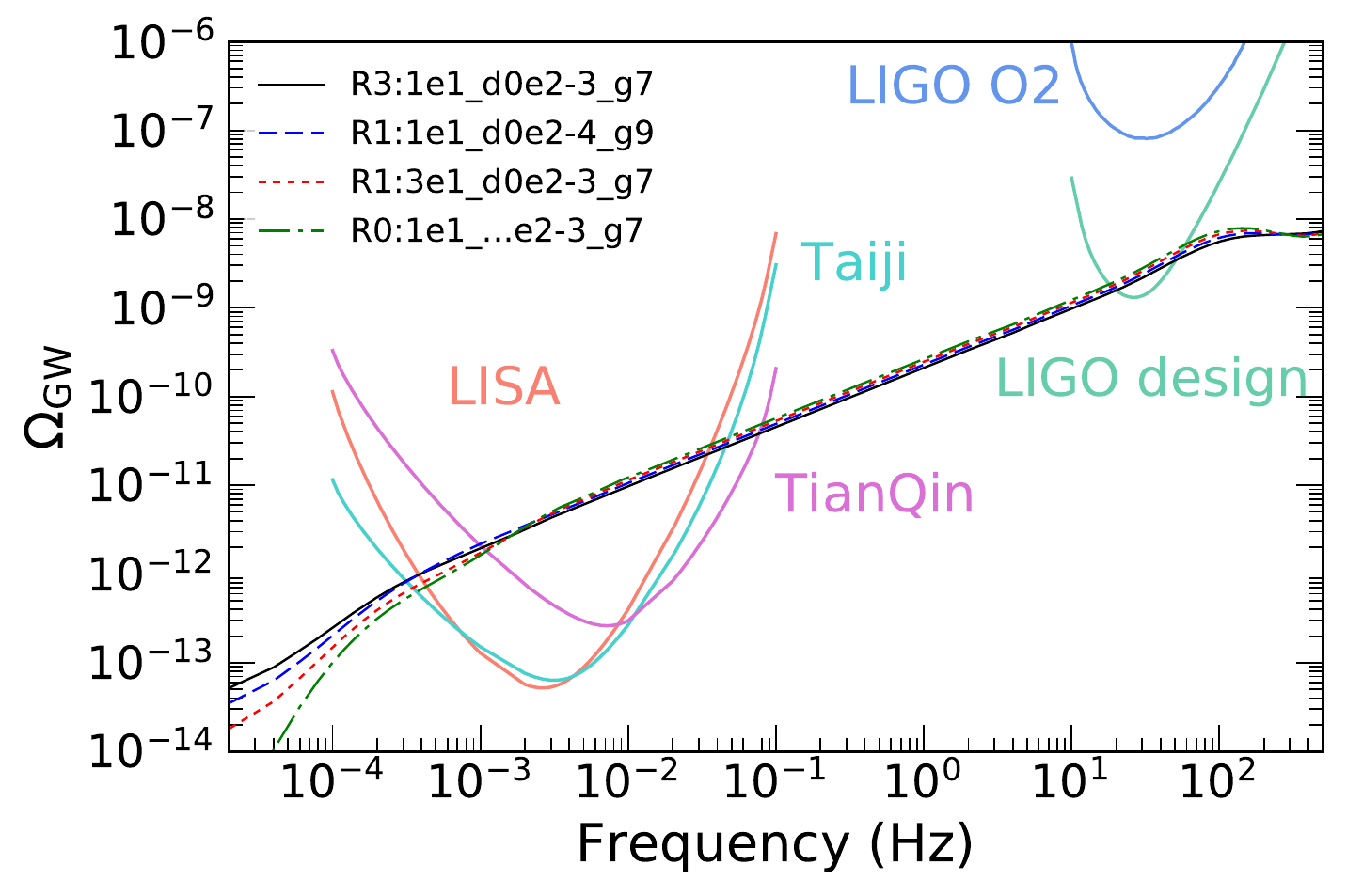}
\caption{
The energy density spectrum of stochastic GWB from sBBHs and BNSs in four different models. In all these models, the EMBS channel sBBHs are circular. The black solid line is from {\bf R3:1e1\_d0e2-3\_g7} model also shown in figure \ref{fig:f2}. Blue dashed line shows the result of {\bf R1:1e1\_d0e2-4\_g9} model where the EMBS channel sBBHs and dynamical channel sBBHs take up the same fraction. Under this model the eccentricities of dynamical sBBHs follow a Gaussian distribution $N(0.9,0.1^2)$ at $10^{-4} \rm Hz$. Red dotted line shows the result from dynamical origin sBBHs dominated model {\bf R1:3e1\_d0e2-3\_g7} where dynamical sBBHs take up $75\%$ and with a Gaussian eccentricity distribution $N(0.7,0.1^2)$ at $10^{-3} Hz$. Green dash dotted line represent a model purely consisted by dynamical origin sBBHs with a high eccentricity distribution($N(0.7,0.1^2)$ at $10^{-3}$ Hz). The sensitivity curves for LISA (red curve), Taiji (cyan curve), TianQin (violet curve), DECIGO (gold curve), LIGO observing runs O2 (blue curve) and LIGO design sensitivity (green curve) are also shown on the figure.
}
\label{fig:f3}
\end{center}
\end{figure}

We calculate the stochastic GWB resulting from the inspiralling and merging of those sBBHs and BNSs in both LISA/Taiji/TianQin and LIGO band for each model (listed in Table~\ref{tab:model}) according to the descriptions in Section~\ref{sec:Ogw} on distributions of sBBH/BNS properties and merger rate density evolution.

Figure~\ref{fig:f2} shows the energy density spectrum of GWB from sBBHs and BNSs resulting from the second to fifth models listed above (see Table~\ref{tab:model}), for which the EMBS channel dominates the formation of sBBHs. It is clear that the GWB signal of inspiralling compact binaries will be detected by both LISA and LIGO with design sensitivity. Top panel of this figure shows the contributions by sBBHs originated from the EMBS and dynamical channel, respectively, and that from BNSs, in {\bf R3:1e1\_d0e2-4\_g9} model. As seen from this panel, the contributions from sBBHs (blue solid line) and BNSs (orange solid line) to the total $\Omega_{\rm GW}$ (black solid line) are more or less the same at most frequency range except at frequency $\lesssim 10^{-3}$\,Hz or $\gtrsim 200$\,Hz, where the BNS contribution becomes more dominant. At low frequencies covered by LISA/Taiji/TianQin,
EMBS channel sources (cyan dashed line) contribute slightly more to the energy density spectrum than those from the dynamical channel as it accounts for three quarter of all the sBBHs.
The large eccentricities of dynamical sBBH sources lead to a rapid drop of their contribution to the energy density spectrum at frequency below $10^{-3}$\,Hz (green dashed line in the top panel). However, this cannot be seen clearly in the total spectrum since the contributions from other sources dominate. The bottom panel of Figure~\ref{fig:f2} shows the total GW energy density spectra of all the four models and their differences in the LISA/Taiji/TianQin band can hardly be seen, and therefore it is difficult to discern these four models simply from the shape and amplitude of $\Omega_{\rm GW}$ (also see Table~\ref{tab:model}).

\begin{figure*}
\begin{center}
\includegraphics[scale=0.65]{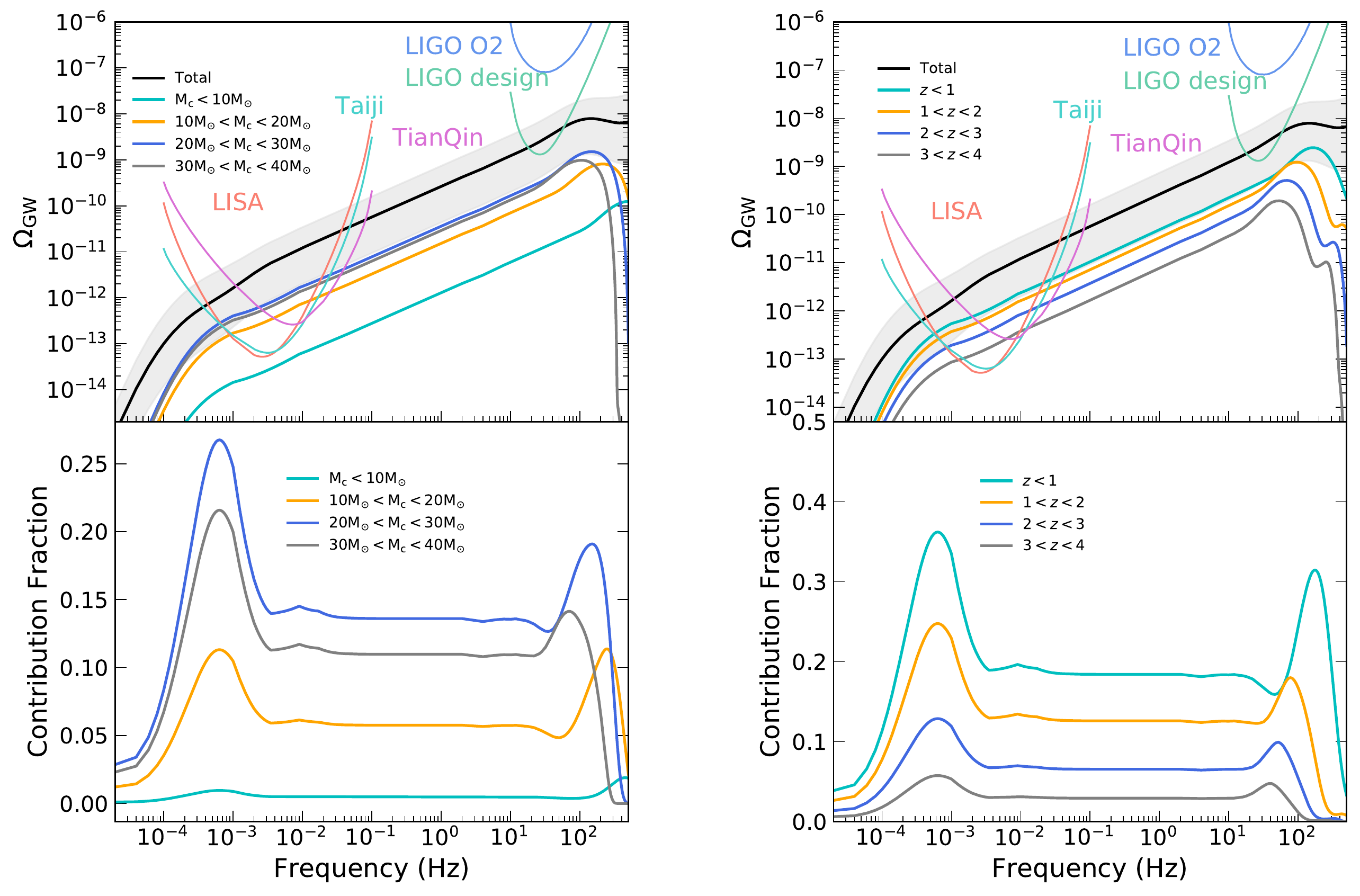}
\caption{
Contributions from sBBH sources within different chirp mass and redshift ranges. The results shown here is for model {\bf R0:1e1\_...e2-4\_g9}: sBBHs are all from dynamical channel with a Gaussian distribution $N(0.9,0.1^2)$ of $e$ at $10^{-4}$ Hz. Top left panel shows the results for sBBHs with different $\mc$ ranges. Bottom left panel shows the contribution fraction relative to the total $\Omega_{\rm GW}$ by adding the BNSs' part. The contributions from sources with $\mc$ between $20 $ and $30 M_{\odot}$ (blue line) and between $30 $ and $40  M_{\odot}$ (gray line) are comparable. sBBHs with $\mc \lesssim 10 M_{\odot}$ contribute only a little to the total GWB (cyan line). Top right and bottom panels show $\Omega_{\rm GW}$ from those sBBHs in different redshift ranges and their contribution fractions with relative to the total GWB. Red, cyan, violet, gold, blue, and green curves represent the sensitivity curves of LISA, Taiji, TianQin, LIGO observing runs O2, and LIGO design sensitivity curves.
}
\label{fig:f4}
\end{center}
\end{figure*} 

Figure~\ref{fig:f3} shows the total energy density spectrum for other three models, i.e., {\bf R1:1e1\_d0-e2-4\_g9}, {\bf R1:3e1\_d0-e2-3\_g7}, and {\bf R0:1e1\_$\cdots$-e2-3\_g7}. Comparing with {\bf R3:1e1\_d0e2-3\_g7} shown also in Figure~\ref{fig:f2}, these three models generate a GWB with larger amplitude and a clear turnover in the shape of $\Omega_{\rm GW}$. The strongest GWB signal is given by {\bf R0:1e1\_...e2-3\_g7} (green dash-dotted line), in which all sBBHs are produced through the dynamical channel and they have relative higher eccentricities following a Gaussian distribution $\rm N(0.7,0.1^2)$ at $10^{-3}$\,Hz. In {\bf R1:1e1\_d0e2-4\_g9} and {\bf R1:3e1\_d0e2-3\_g7}, it is assumed that half and a quarter of sBBHs are originated from the EMBS channel, thus the GWB produced by them lie in between the other two models in Figure~\ref{fig:f3} at frequencies $\gtrsim 0.002$ Hz. The amplitude differences largely depend on the fraction of sBBHs contributed by different formation channels. The larger the contribution fraction of dynamical sBBHs is, the stronger the GWB becomes because the dynamical sBBHs have relatively larger chirp masses. 

We also list the resulting GWB energy density at $25\rm{Hz}$ for all our models in Table~\ref{tab:model}. The values range from $1.68 \times 10^{-9}$ to $2.38 \times 10^{-9}$, and their difference is small, at most a factor of $\sim 1.4$. In {\bf R1:0e1\_d0e2\_...}, our estimate is smaller than the value $1.8\times 10^{-9}$ given by \citet{2018PhRvL.120i1101A} because of a smaller local merger rate density adopted in the present work. For some models listed in Table~\ref{tab:model}, we obtain higher $\Omega_{25\rm Hz}$ because of significant contribution from the dynamically originated sBBHs, with high relatively chirp masses.

Apart from the amplitude of $\Omega_{\rm GW}(f)$, the shape of GW energy density spectrum also contains the information for the origin of these compact binaries. Eccentricity distribution will influence the shape of $\Omega_{\rm GW}(f)$ causing a drop at low frequency. As seen from Figure~\ref{fig:f2}, the shape of $\Omega_{\rm GW}(f)$ changes little if the eccentricities of sBBHs are not significant as the model set. If the eccentricities of sBBHs are large at high frequency (e.g., $0.9$ at $\gtrsim 10^{-3}$\,Hz), the shape of $\Omega_{\rm GW}(f)$ is significantly bent in the LISA band (see Fig.~\ref{fig:f3}). In order to quantify the influence of eccentric binaries on GWB energy density spectrum, we fit $\Omega_{\rm GW}(f)$ generated by all the models by a double power-law form to account for the shape bending as
\begin{equation}
    \label{eq:fit}
    \Omega_{\rm GW}(f) = A\times
    \begin{cases}
    (f/f_*)^{\frac{2}{3}}, & \textrm{if}\ f_* < f < 1\,{\textrm{Hz}} \\
    (f/f_*)^{\beta}, & \textrm{if}\ f < f_*, 
    \end{cases}
\end{equation}
where $A$ is the amplitude at bending frequency $f_*$, $\beta$ is the power index at low-frequency. Our fitting results are listed in Table~\ref{tab:model}. For the high-frequency part, the power index resulting from each model is fixed at the unique value of $2/3$ because most, if not all, sBBHS are circularized due to GW radiation. However, for the low-frequency part, the power index $\beta$ resulting from those models with dynamical sBBHs dominating the contribution to GWB can be substantially larger than the unique value $2/3$ because of their high eccentricity. 

If the EMBS sources are the main contributors, then $\beta$ differs little from $2/3$ because of the small contribution from dynamical sBBHs with high eccentricities. The largest difference between $\beta$ and $2/3$ is obtained from {\bf R0:1e1\_...e2-3\_g7}, in which sBBHs are all formed from the dynamical channel with relatively the largest eccentricities among all the models considered here. The differences on the resulting shape and amplitude of $\Omega_{\rm GW}(f)$ from different models suggest that it is possible to discern various sBBH formation channels by accurately measuring the GWB with LISA. 
According to our fitting results listed in Table~\ref{tab:model}, we note here that the turnover frequency $f_*$ may be below the LISA sensitive band (a few times $10^{-4}$ Hz) in several models, therefore it may be difficult to be observationally determined. However if the eccentricities of sBBHs could be relatively larger at higher frequencies than those assumed in the present paper, the turnover point may move to higher frequency and thus the double power-law shape of GWB spectrum may be easier to be observed. As mentioned in Section~\ref{sec:basic}, sBBHs formed from the AGN-assisted channel or sBBHs inspirals and mergers induced by the Lidov-Kozai mechanisms, ignored in our models, can have relatively higher eccentricities than what we assumed in the present paper, and thus may help to lead to a higher turnover frequency.
We also note here that that the effect from eccentric sBBHs on the GWB spectrum is constrained by the contribution from cosmic BNSs. If the real local BNS merger rate density is substantially smaller than the current constraint from GW observations, the value of $\beta$ for each model would become larger than that listed in Table~\ref{tab:model}.

Figure~\ref{fig:f4} shows the contributions to GWB by sBBHs with different properties or at different redshift range for the model {\bf R0:1e1\_...e2-4\_g9} as an example. (Results from other models are similar to this one.) From the left two panels of Figure~\ref{fig:f4}, we can see that sources with chirp mass $\mathcal{M}_{\rm c} \lesssim 10 M_{\odot}$ contribute little to the total GWB, while those with $\mathcal{M}_{\rm c} \sim 20-30 M_{\odot}$ contribute more or less the same to GWB as those with $\mathcal{M}_{\rm c} \sim 30-40 M_{\odot}$, except at the high- and low-frequency ends, and the sources with $\mathcal{M}_{\rm c} \sim 10-20 M_{\odot}$ contribute less than those with $\mathcal{M}_{\rm c} \sim 20-30 M_{\odot}$.
For sources with higher $\mathcal{M}_{\rm c}$, their contribution to GWB becomes small because their merger rate density declines rapidly with increasing $\mathcal{M}_{\rm c}$. Right panels of Figure~\ref{fig:f4} show the contributions from sBBHs at different redshift ranges. The relative contribution fraction of those sources at $z<3$ compared with all sBBHs is more than $90\%$. The shift of the peaks at the high-frequency end is due to that the GWB from sBBHs with different chirp masses drop at different frequencies.
At both the low- and high-frequency ends, the contribution from BNSs becomes dominant (see top panel of Fig.\ref{fig:f2}). At the high-frequency end ($\gtrsim 300$\,Hz), the contribution from sBBHs drops because they only emit GWs at lower frequencies, while at the low frequency end ($\lesssim 10^{-4}$\,Hz), it drops dramatically in some models because of the high eccentricities. 
As seen from the bottom panels, the contributions from sBBHs in different mass ranges have a peak at both the high-frequency and low-frequency ends. At the high-frequency end, the peaks are caused by the maximum GW radiation and subsequent rapid drop of sBBHs at the merger phase, where the contribution from DNSs is significant and does not decline. The frequency of the peak increases with decreasing sBBH mass range as the GWB spectrum from lighter sBBHs drops at lower frequency. At the low-frequency end, the peak is caused by the effect of eccentricity. Eccentric sBBHs emit GW energy most at their peak frequencies but has little power at lower frequencies, so the contribution rises at the frequency (slight) higher than $10^{-4}$\,Hz, where we assume that dynamically originated sBBHs have high eccentricities.

\subsubsection{Signal-to-Noise ratio (SNR)}
\label{sec:snr}

The GWB from sBBHs and BNSs estimated above may be detected by LISA/Taiji/TianQin and LIGO/Virgo/KAGRA. The signal-to-noise ratio (SNR) for the predicted GWB resulting from each model can be estimated according to the sensitivity curves of those detectors. 

The expected SNR of the GWB ($\Omega_{\rm GW}$), if detected by LISA, can be estimated as \citep{2013PhRvD..88l4032T}
\be 
\label{eq:snrlisa}
{\rm SNR}=\sqrt{T}\left[\int_{0}^{\infty}\frac{\Omega_{\rm GW}^2(f)}{\Omega_{\rm n}^2(f)} df\right]^{1/2}.
\ee
Here $\Omega_{\rm n}(f) = \frac{2 \pi^2 f^3 S_{\rm n}(f)}{3 H_0^2}$, $S_{\rm n}(f) = \frac{P_{\rm n}(f)}{\mathcal{R}(f)}$ is the strain spectral sensitivity, $\mathcal{R}(f)$ is the transfer function of the detector,  $P_{\rm n}(f)$ is its noise power spectral density, $T$ is the total observation time and set as $T=5$\,years the same as that adopted by \citet{2016PhRvL.116w1102S}. 
We adopt the following fitting formula for $\mathcal{R}(f)$ of LISA given by \citep{2018arXiv180301944C}
\be
\mathcal{R}(f) = \frac{3}{10}\frac{1}{1+0.6(f/f_*)^2},
\ee
where $f_* = 1.909\times 10^{-2}$\,Hz. 

The expected SNR of the GWB ($\Omega_{\rm GW}$), if detected by LIGO, can be roughly estimated as \citep{1993PhRvD..48.2389F, 2018PhRvL.120i1101A}
\be
{\rm SNR}=\frac{3H_0^2}{10\pi^2}\sqrt{2T}\left[ \int_{0}^{\infty} df \sum_{i>j}
\frac{\gamma_{ij}^2(f)\Omega_{\rm GW}^2(f)}{f^6P_{\rm n}^i(f)P_{\rm n}^j(f)} \right]^{1/2},
\label{eq:eqsnr}
\ee
where $\gamma_{ij}$ is the overlap reduction function \citep{1993PhRvD..48.2389F}, $P_{\rm n}^i$ and $P_{\rm n}^j$ are the noise power spectral densities in the two detectors, $T$ is also the duration of observation(s). In the present paper we adopt $T=24$\,months. 

We present SNR results of all our models in Table~\ref{tab:snr}. The differences between four EMBS channel dominated models in Table~\ref{tab:model} are quite small. The largest SNR result is from the model {\bf R0:1e1\_...e2-3\_g7}, in which all the sBBHs are originated from dynamical channel. The expected SNRs for LISA are quiet large for all those models, while they are much smaller for LIGO. For different models, the expected SNR differ a lot in both types of detectors because of the differences in the amplitude of the predicted $\Omega_{\rm GW}$. The higher the fraction of dynamical sBBHs, the higher the amplitude of the GWB, and thus the larger the expected SNR.

\begin{table*}
\centering
\caption{Expected SNRs of the predicted GWB from different models if detected by LISA, Taiji, TianQin and LIGO detectors with design sensitivities. }
\label{tab:snr}
\begin{tabular}{|c|cc|cc|cc|cc|} 
\hline
\multirow{2}{*}{Model} & \multicolumn{2}{c}{LISA} & \multicolumn{2}{c}{Taiji} & \multicolumn{2}{c}{TianQin} & \multicolumn{2}{c}{LIGO} \\ \cline{2-3} \cline{4-5} \cline{6-7} \cline{8-9}
        & sBBH & Total & sBBH &  Total & sBBH & Total & sBBH & Total  \\   \hline
{\bf R1:0e1\_d0e2\_...} & $129^{+121}_{-104}$ &   $274^{+584}_{-232}$ & $120^{+114}_{-97}$  & $255^{+542}_{-216}$ & $10.1^{+9.5}_{-8.16}$  & $21.2^{+44.9}_{-17.9}$ & $1.47^{+1.39}_{-1.19}$   & $2.88^{+5.88}_{-2.43}$ \\ \hline
{\bf R3:1e1\_d0e2-4\_g9} & $150^{+142}_{-121}$ &   $296^{+604}_{-250}$ & $141^{+133}_{-114}$  & $276^{+561}_{-233}$ & $12.0^{+11.3}_{-9.69}$  & $23.1^{+46.7}_{-19.5}$ & $1.78^{+1.68}_{-1.44}$   & $3.19^{+6.16}_{-2.68}$ \\ \hline
{\bf R3:1e1\_g3e2-4\_g9} & $142^{+133}_{-115}$ &   $287^{+596}_{-242}$ & $134^{+125}_{-108}$  & $268^{+554}_{-226}$ & $11.8^{+11.2}_{-9.53}$  & $23.0^{+46.4}_{-19.4}$ & $1.78^{+1.68}_{-1.44}$   & $3.19^{+6.16}_{-2.68}$ \\ \hline
{\bf R3:1e1\_d0e2-3\_g7} & $145^{+136}_{-117}$ &   $290^{+598}_{-245}$ & $137^{+130}_{-111}$  & $272^{+557}_{-229}$ & $12.0^{+11.3}_{-9.70}$  & $23.1^{+46.6}_{-19.5}$ & $1.78^{+1.68}_{-1.44}$   & $3.19^{+6.16}_{-2.68}$ \\ \hline
{\bf R3:1e1\_g3e2-3\_g7} & $136^{+128}_{-110}$ &   $281^{+590}_{-237}$ & $130^{+122}_{-105}$  & $264^{+551}_{-223}$ & $11.8^{+11.1}_{-9.53}$  & $22.9^{+46.5}_{-19.3}$ & $1.78^{+1.68}_{-1.44}$   & $3.19^{+6.16}_{-2.68}$ \\ \hline
{\bf R1:1e1\_d0e2-4\_g9} & $172^{+162}_{-139}$ &   $318^{+624}_{-268}$ & $162^{+152}_{-131}$  & $296^{+581}_{-249}$ & $13.9^{+13.0}_{-11.2}$  & $25.0^{+48.4}_{-21.0}$ & $2.09^{+1.97}_{-1.69}$   & $3.50^{+6.45}_{-2.93}$ \\ \hline
{\bf R1:1e1\_d0e2-3\_g7} & $161^{+151}_{-130}$ &   $306^{+613}_{-258}$ & $154^{+146}_{-124}$  & $289^{+573}_{-243}$ & $13.8^{+13.0}_{-11.2}$  & $25.0^{+48.3}_{-21.0}$ & $2.09^{+1.97}_{-1.69}$   & $3.50^{+6.45}_{-2.93}$ \\ \hline
{\bf R1:3e1\_d0e2-4\_g9} & $194^{+182}_{-157}$ &   $339^{+645}_{-284}$ & $182^{+172}_{-147}$  & $317^{+600}_{-266}$ & $15.8^{+14.8}_{-12.8}$  & $26.9^{+50.2}_{-22.5}$ & $2.40^{+2.25}_{-1.94}$   & $3.81^{+6.74}_{-3.18}$ \\ \hline
{\bf R1:3e1\_d0e2-3\_g7} & $177^{+166}_{-143}$ &   $322^{+627}_{-271}$ & $172^{+161}_{-139}$  & $306^{+589}_{-257}$ & $15.7^{+14.7}_{-12.7}$  & $26.8^{+50.1}_{-22.5}$ & $2.40^{+2.25}_{-1.94}$   & $3.81^{+6.74}_{-3.18}$ \\ \hline
{\bf R0:1e1\_d0e2-4\_g9} & $216^{+203}_{-175}$ &   $361^{+665}_{-302}$ & $203^{+191}_{-164}$  & $338^{+619}_{-283}$ & $17.6^{+16.6}_{-14.2}$  & $28.8^{+51.9}_{-24.1}$ & $2.70^{+2.55}_{-2.18}$   & $4.12^{+7.03}_{-3.43}$ \\ \hline
{\bf R0:1e1\_...e2-3\_g7} & $193^{+182}_{-156}$ &   $338^{+642}_{-284}$ & $189^{+178}_{-150}$  & $323^{+605}_{-271}$ & $17.5^{+16.5}_{-14.1}$  & $28.7^{+51.8}_{-24.0}$ & $2.70^{+2.55}_{-2.18}$   & $4.12^{+7.03}_{-3.43}$ \\ \hline
{\bf R0:1e1\_...e2-3\_U(0.5,1)} & $207^{+196}_{-167}$ & $353^{+656}_{-296}$ & $198^{+187}_{-160}$ & $333^{+614}_{-279}$ & $17.6^{+16.6}_{-14.2}$ & $28.7^{+16.6}_{-14.2}$ & $2.70^{+2.55}_{-2.18}$   & $4.12^{+7.03}_{-3.43}$ \\ \hline
\end{tabular}
\begin{flushleft}
\footnotesize{Note: the observations time period for LISA (Taiji, TianQin) and LIGO are set as $5$\,years and $24$\,months, respectively. First column denotes the model name. Second, third, and fourth columns show the expected SNR  values for the predicted GWB from sBBH, BNS, and all sources by LISA. The last three columns correspond to those by LIGO.}
\end{flushleft}
\end{table*}

\section{Simulating the GWB Signal in the time domain}
\label{sec:GWsignal}

We also estimate the GWB signal in the time domain that may be detected by LISA and LIGO/VIRGO/KAGRA, in addition to the GWB energy density spectrum estimated above. In Section~\ref{sec:sample}, we first describe how to generate mock samples of sBBHs and BNSs, which can be used to obtain the GWB signal by direct summation of the GW radiation from different sources across the cosmic time. In section~\ref{sec:signal}, we present the simulated GWB time serious signal in the LISA bands. For the simulation of time domain GWB signals in the LIGO/VIRGO/KAGRA band, see \citet{2018PhRvL.120i1101A}. 

\subsection{Mock Samples}
\label{sec:sample}

The total number density of GW sources at an orbital period range from $P$ to $P+dP$ can be estimated if the merger rate density evolution ${\Rmrg(\mc,e_0,z)}$ of these sources with initial eccentricities $e_0$ at an orbital period $P_{0}$ is known. In the present paper, we choose $P_{0}=10^{5}$\,s to study the number distribution of GW sources in both LISA and LIGO bands. Assuming all binaries are circular ($e_0=0$) when they emit GWs in the LISA and LIGO/VIRGO/KAGRA bands, the total number of circular sources in the inspiral stage can be calculated by
\begin{equation}
\frac{dN}{dP} \simeq \frac{dN}{dt}\cdot \frac{dt}{dP} = 
\iint \frac{5\Rmrg(\mc,0,z)c^5  P^{5/3}}{384 \cdot 2^{2/3}\pi^{8/3}(G\mc)^{5/3}} \frac{dV}{dz} dz d\mc.
\label{eq:num-circ}
\end{equation}
We also consider the merger and ringdown stages according to the analytic fit to the GW energy spectrum given in Equation~(\ref{eq:gw-circ}). If the GW sources in the inspiral stage are non-circular with an eccentricity distribution of $P(e_0)$ at $P_0$, the above Equation~(\ref{eq:num-circ}) can be modified to 
\begin{eqnarray}
    \frac{dN}{dP} \simeq \iiint 
    \frac{5\bm{\Rmrg}(\mc,e_0,z)c^5(1-e^2)^{7/2}  P^{5/3}}{384\cdot 2^{2/3} \pi^{8/3} (G\mc)^{\frac{5}{3}}
    \left(1+\frac{73}{24}e^2+\frac{37}{96}e^4\right)} \nonumber\\
 	\times \frac{dV}{dz} dz d\mc d e_0 ,
\label{eq:num-ecc}
\end{eqnarray}
where $e$ is given by Equation~(\ref{eq:f2e}) as $e=e(f_{\rm p}; e_0, f_{\rm p,0})$ with $f_{\rm p}=1/P$. 

\begin{figure}
\centering
\includegraphics[scale=0.6]{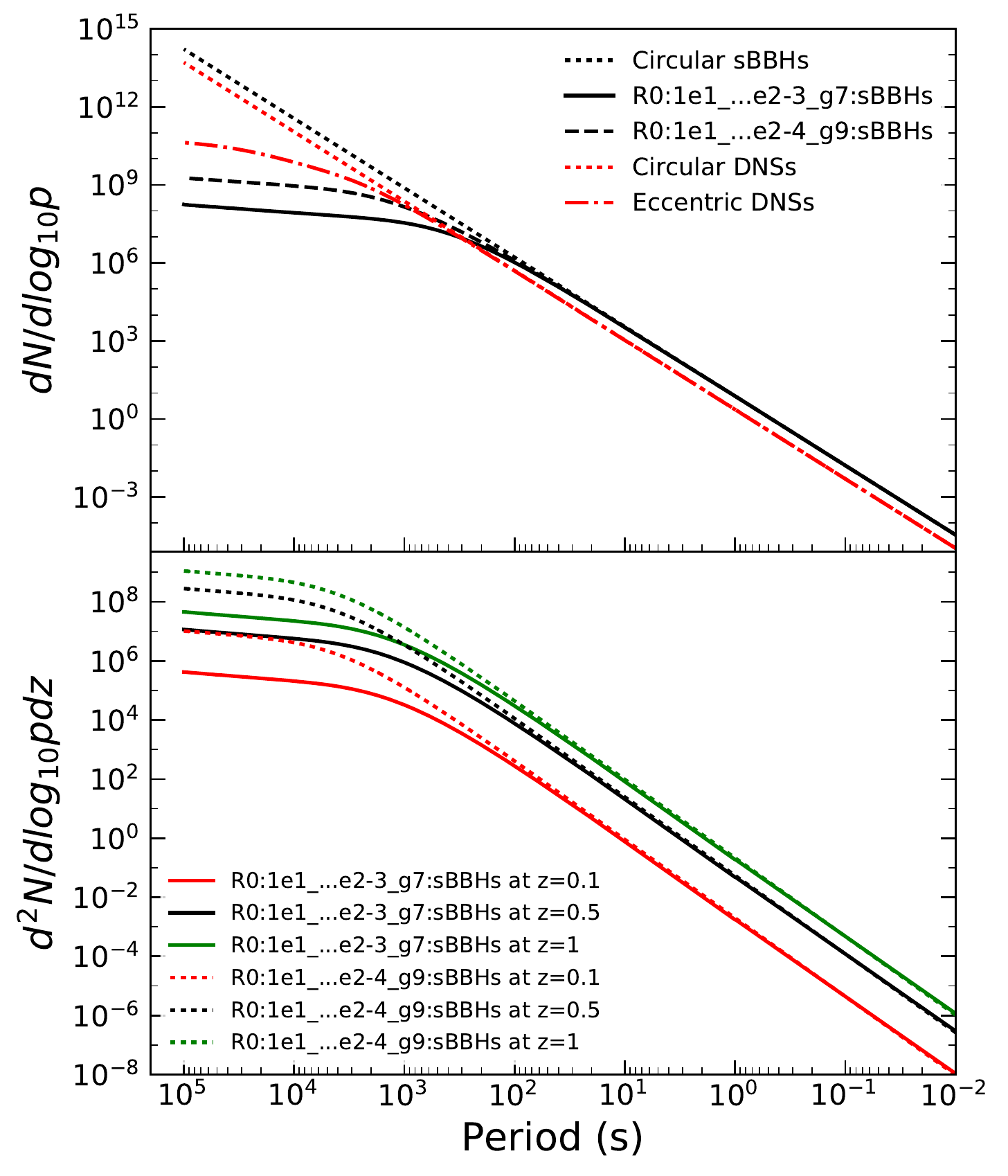}
\caption{
Period distributions for GW sources of different models. In the top panel, black and red dotted lines show the circular cases for sBBHs and BNSs respectively. Black solid line indicates the period distribution of sBBHs in {\bf R0:1e1\_...e2-3\_g7} model, black dashed line indicates that in {\bf R0:1e1\_...e2-4\_g9} model and red dash-dotted line indicates eccentric BNSs used in all above models with eccentricity distribution $N(0.7,0.1^2)$ at $10^{-4}$ Hz. In the bottom panel, we show period distribution at different redshift for two models: {\bf R0:1e1\_...e2-3\_g7} (solid lines) and {\bf R0:1e1\_...e2-4\_g9} (dotted lines).
}
\label{fig:f5}
\end{figure}

Figure \ref{fig:f5} shows the results of the period distribution of GW sources from two different models: {\bf R0:1e1\_...e2-3\_g7} and {\bf R0:1e1\_...e2-4\_g7}. 
In the top panel, black and red dotted lines show $dN/d\log_{10}P$ from circular sBBHs and BNSs respectively. Circular cases show a single power law relation as Equation (\ref{eq:num-circ}) indicates. Compared with circular cases, period distribution of initially eccentric sources drops dramatically at long period range because these sBBHs radiate GW more rapidly and decay to shorter period faster than circular ones. For different eccentricity distribution the turning point is different, {\bf R0:1e1\_...e2-3\_g7} model shows deviation from circular case at the largest $P$. In the bottom panel of Figure~\ref{fig:f5}, we show the differential distribution $d^2N/d\log_{10}Pdz$ at different redshift obtained from the above two models.

From period distribution we can get GW source number in our mock sample. However when producing the mock samples for sBBHs and BNSs, we only pick those with orbital frequency between $0.004$\,Hz and $0.5$\,Hz for the following reasons. First, the number of compact binaries at lower frequencies are too many to be efficiently calculated. Second, the contribution to the total power from sources at lower frequencies is much less significant than those from higher frequencies. 
Within this frequency range, we ignore the effect of eccentricity as almost all sources have been circularized (see Figure~\ref{fig:f5}). We assign physical parameters ($m_1, q, z$) to each individual sBBH and BNS according to the probability distributions of these parameters. We use two models to demonstrate our results from mock samples. In EMBS dominated model, whose sBBH fraction setting is the same as the second to fifth models in Table \ref{tab:model}, sBBHs are composed by both EMBS channel sources and dynamical channel sources whose physical parameter distributions are described clearly in Section~\ref{sec:mr_bbh}. As for the other purely dynamical channel model, whose sBBH fraction setting is the same as last two models in Table \ref{tab:model}, all the sBBHs are from dynamical origin channel and their source parameter distributions are described in Section~\ref{sec:mr_bbh}. The parameter distributions of BNSs are described in Section~\ref{sec:mr_BNS}. Except for these three physical parameters we also need to consider their position information when calculating the strain signal in time domain. The directions of the sBBHs/BNSs orbital planes should be randomly distributed on the sky and thus its orientations with respect to the gravitational wave detector are random distributed. 

\subsection{ GWB signal at the LISA band}
\label{sec:signal}

From the mock sample of sBBHs and BNSs, we can directly estimate their GWB strain signal in the time domain.
In the LISA band ($10^{-4}$-$1$\,Hz), all sBBHs/BNSs are radiating GWs in the inspiral phase with semimajor axes ($a \gg GM_{\rm tot}/c^2$), and thus the Newtonian approximation of the GW strain should be sufficiently accurate for the estimation of the GWB signal. For a binary system with chirp mass $\mc$, distance $r$ and inclination angle $\iota$ between the line of sight (LOS) and the orbital angular momentum direction of the binary system, the two polarizations of GW are given by:
\begin{eqnarray}
h_+(t) & = & - \frac{G \mc}{c^2 r}\frac{1+\cos^2 \iota}{2} 
\left( \frac{c^3(t_{\rm c}-t)}{5G\mc} \right)^{-1/4}\times \nonumber \\
& & 
\cos\left[2\varphi_{\rm c} - 2\left(\frac{c^3(t_{\rm c}-t)}{5G\mc}\right)^{5/8} \right],
\end{eqnarray} 
\begin{eqnarray}
h_{\times}(t)= - \frac{G \mc}{c^2 r} \cos \iota \left( \frac{c^3(t_{\rm c}-t)}{5G\mc} \right) ^{-1/4}\times \nonumber \\
\sin\left[2\varphi_{\rm c} - 2\left(\frac{c^3(t_{\rm c}-t)}{5G\mc}\right)^{5/8} \right].
\end{eqnarray} 
Here $\varphi_{\rm c}$ is the phase of GW at the merger time $t_{\rm c}$ (i.e., the time of coalescence). To account for the orientations of GW detectors, the antenna pattern functions for LISA-like detectors:
\be
F_{+}=\frac{\sqrt{3}}{2}\left[\frac{1}{2}(1+\cos^2\theta)\cos2\phi 
\cos2\psi-\cos\theta \sin2\phi \sin2\psi \right],
\label{eq:eq22}
\ee
and
\be
F_{\times}=\frac{\sqrt{3}}{2}\left[\frac{1}{2}(1+\cos^2\theta)\cos2\phi 
\sin2\psi+\cos\theta \sin2\phi \cos2\psi \right],
\label{eq:eq23}
\ee
are required to be convoluted in. Here parameters ($\theta, \phi$) describe the sky location of the source and $\psi$ denotes the orientation of the source relative to the detector. For detailed discussions, see \citet{2003PhRvD..67b2001C}. The strain measured by the detector is then
\be
\label{eq:eq24}
h=F_{+}h_{+}(t)+F_{\times}h_{\times}(t).
\ee
By summing up the strain of all sources from the mock sample together, we can then obtain the simulated GWB strain signal. We calculate a time series of duration $10^{4}$ seconds to present the GWB signal of sBBHs and BNSs \citep[see also][]{2018PhRvL.120i1101A}. During the period of observation time $T_{\rm p}$, we set the sampling rate to $0.3\ {\rm s}^{-1}$, which is about the proposed sampling rate of LISA ($\sim 3$\,Hz; \citealt{2017arXiv170200786A}). 

\begin{figure}
\begin{centering}
\includegraphics[scale=0.6]{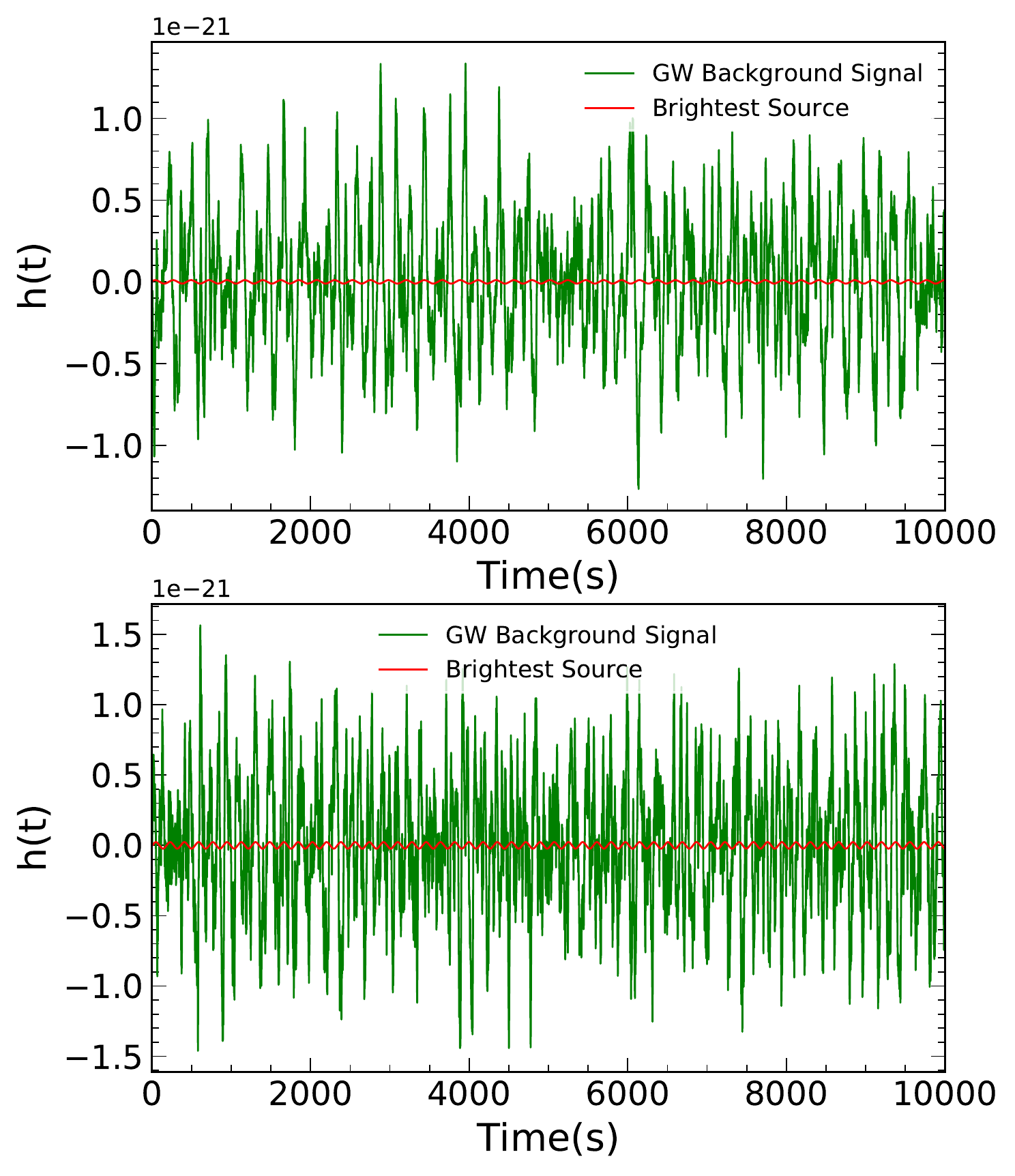}
\caption{Simulated GWB signal from the sBBHs and BNSs in two different models:
{\bf R3:1e1\_d0e2-4\_g9}, where the EMBS dominates the formation of sBBHs (top panel), and {\bf R0:1e1\_...e2-4\_g9}, where all sBBHs are formed via the dynamical channel model (bottom panel). Note that here we ignore the eccentricities of all the sources because it is not important in the frequency range we consider here. The observation time duration shown here is $10^{4}$ s. The signal from each model is on the order of $10^{-21}$. Bottom purely dynamical channel model shows a slightly larger amplitude compared to the EMBS dominated model, which is consistent with the result of $\Omega_{\rm GW}$. The red line shows that the signal from brightest source is much smaller than the GWB signal. For both models the brightest signal is two orders of magnitude smaller than total GWB signal.}
\label{fig:f6}
\end{centering}
\end{figure}

Figure~\ref{fig:f6} shows the sBBHs and BNSs GWB strain signal in the time domain resulting from the EMBS dominated model and purely dynamical model. Here we do not consider the eccentricity, so we simplify the models above in Table~\ref{tab:model}. In the EMBS dominated model, the EMBS channel sBBHs are about $75\%$ of all sBBHs, while in the purely dynamical model all sBBHs are formed via the dynamical channel. 
In this Figure, we only show the signal of the GWB from binary systems at inspiral stage without considering the instrument noise, for simplicity. The instrument noise can be simply modeled by integrating over the LISA noise power spectrum. The GWB strain amplitude is on the order of $10^{-21}$, which should be detectable by LISA. For comparison, the strain amplitude of the brightest source (an sBBH, shown by the red lines in Fig.~\ref{fig:f6}) in the mock sample generated from each model is 
on the order of $10^{-23}$, about two orders of magnitude smaller than the GWB signal.
We also check the BNSs' signal, their strain signals are significantly smaller than those of sBBHs. 
As also seen from Figure~\ref{fig:f6}, the purely dynamical channel model (bottom panel) produces a slightly higher GWB signal than the other model because the sBBHs in this model have larger chirp masses compared to those in the other model. More specifically, $\left<h^2 \right>$ equals $2.3\times10^{-43}$ in purely dynamical channel model while $\left<h^2 \right>$ is  $1.6\times10^{-43}$ for EMBS dominated model. These two numbers are consistent with $\Omega(f)$ estimated previously.

\subsection{Signal-to-Noise Ratio (SNR) for individual sources}
\label{sec:indsource}

We also estimate SNRs for individual sources in our mock sample in the LISA band, in addition to that for the GWB. In this case, the GWB may be considered as unresolved background noises and the effective noise power spectral density can be estimated as $S_{\rm eff}(f) = S_{\rm n}(f)+S_{\rm GWB}(f)$. Here $S_{\rm GWB}$ is given by \citep[see][]{2004PhRvD..70l2002B,2018ApJ...864...61C}:
\begin{equation}
S_{\rm GWB}(f) = \frac{3H_0^2}{2\pi^2}\frac{\Omega_{\rm GW}(f)}{f^3},
\end{equation}
and $S_{\rm n}(f)$ for LISA can be found in Section~\ref{sec:snr}. The SNR for individual sources can then be estimated as  \citep[see][]{2004PhRvD..70l2002B}:
\begin{equation}
({\rm SNR})^2 = 2\int \frac{2\pi}{3(\pi D)^2c^3}\frac{(G\mc)^{5/3}(\pi f)^{-1/3}}{(1+z)f^2S_{\rm eff}(f)}df.
\end{equation}
Here we adopt a 5-year mission time for LISA,
which is reflected in the exact frequency range we integrate in the above formula for each source. 
Since the GWB signal acts as a stochastic noise, the expected SNRs for individual sources are relatively lower compared with those without considering such a noise. 
If the GWB can be modelled well, the SNR for individual sources may be also obtained by removing the GWB, in which case the noise power spectral density is $S_{\rm n}(f)$. In this case, we also estimate the expected SNRs for individual sources. 
Here we pick those mock sources with ${\rm SNR} > 8$ as ``detectable'' objects. Apparently BNSs cannot be detected by LISA (usually with SNR $\lesssim 2$). Note that we only generate one mock sample of BNSs as the formation recipes for BNSs are the same in different models considered in this paper. 

Considering the uncertainties from intrinsic merger rate density, we find that the numbers of detection for EMBS dominated model and purely dynamical channel model are about $94^{+90}_{-89}$ and $110^{+111}_{-98}$, respectively. Considering of the confusion noise from GWB, these numbers decrease to $81^{+44}_{-76}$ and $85^{+51}_{-73}$, respectively. If the mission time for LISA extends to $10$\,years, the corresponding numbers for these two models are about $267^{+272}_{-253}$ and $265^{+266}_{-241}$ when confusion noise of GWB is removed. While treating GWB as unknown noise, those numbers are reduced to $228^{+128}_{-214}$ and $217^{+138}_{-194}$, respectively. 
It is clear that the expected SNR of an sBBH when considering the GWB noise is smaller, but not much smaller, than that without considering such noise, and the expected number of the ``detectable'' mock sBBHs does change somewhat.
In the generation of mock objects via the Monte-Carlo method, only those systems with $f>0.004$\,Hz are generated but GW sources at lower frequencies are excluded. There might be some sources with lower frequencies that may be missed in the above estimates. 

\begin{figure*}
\centering
\includegraphics[scale=0.6]{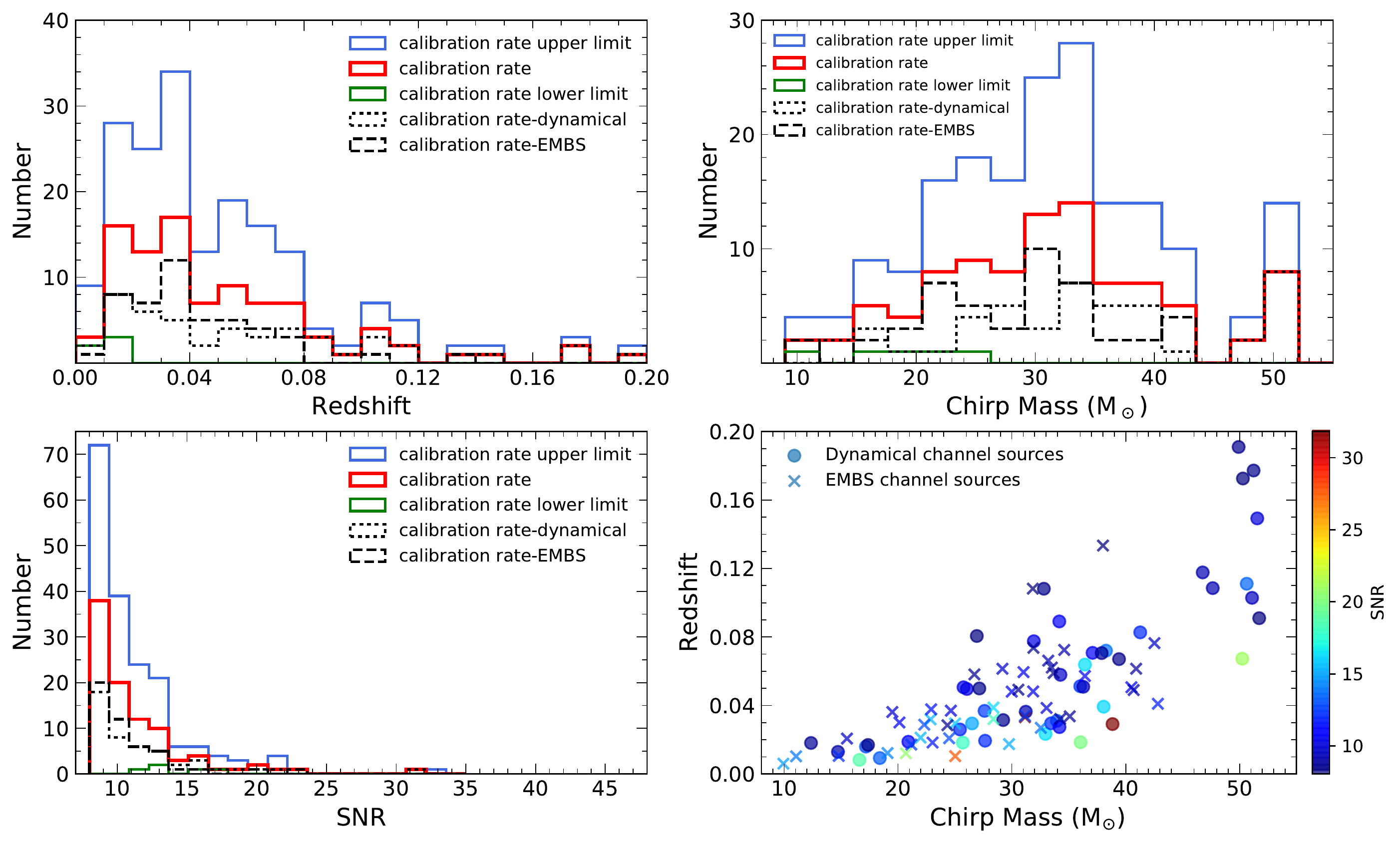}
\caption{Distributions of the properties of the ``detectable'' mock BBHs (with $\mathrm{SNR} > 8$ by $5$ years observation of LISA) resulting from the EMBS dominated model. Top-left, top-right, and bottom-left panels show the redshift distribution, the chirp mass distribution, and the expected SNR distribution, respectively. In each of these three panels, the red histogram shows the distribution of ``detectable'' mock BBHs obtained by using the BBH merger rate density calibrated by the observational constraint (mean value) from LIGO/VIRGO, and the dotted histogram and the dashed histogram separately show the distributions of those BBHs from the EMBS channel and the dynamical channel, respectively. The blue and green histograms represent the distributions of ``detectable'' mock BBHs by considering the uncertainty with $90\%$ confidence level of the constraint on the merger rate density from LIGO, denoted as ``calibration rate upper limit'' and ``calibration rate lower limit", respectively.
The bottom-right panel shows the distribution of these ``detectable'' mock BBHs both from the EMBS channel (crosses) and the dynamical channel (filled circles) on the chirp mass-redshift plane with SNR indicated by the color indexes, and in this panel, only the results obtained by using the BBH merger rate density calibrated by the LIGO/VIRGO observations, but without considering the uncertainty of the constraint on the BBH merger rate density.
}
\label{fig:f7}
\end{figure*}

\begin{figure*}
\centering
\includegraphics[scale=0.6]{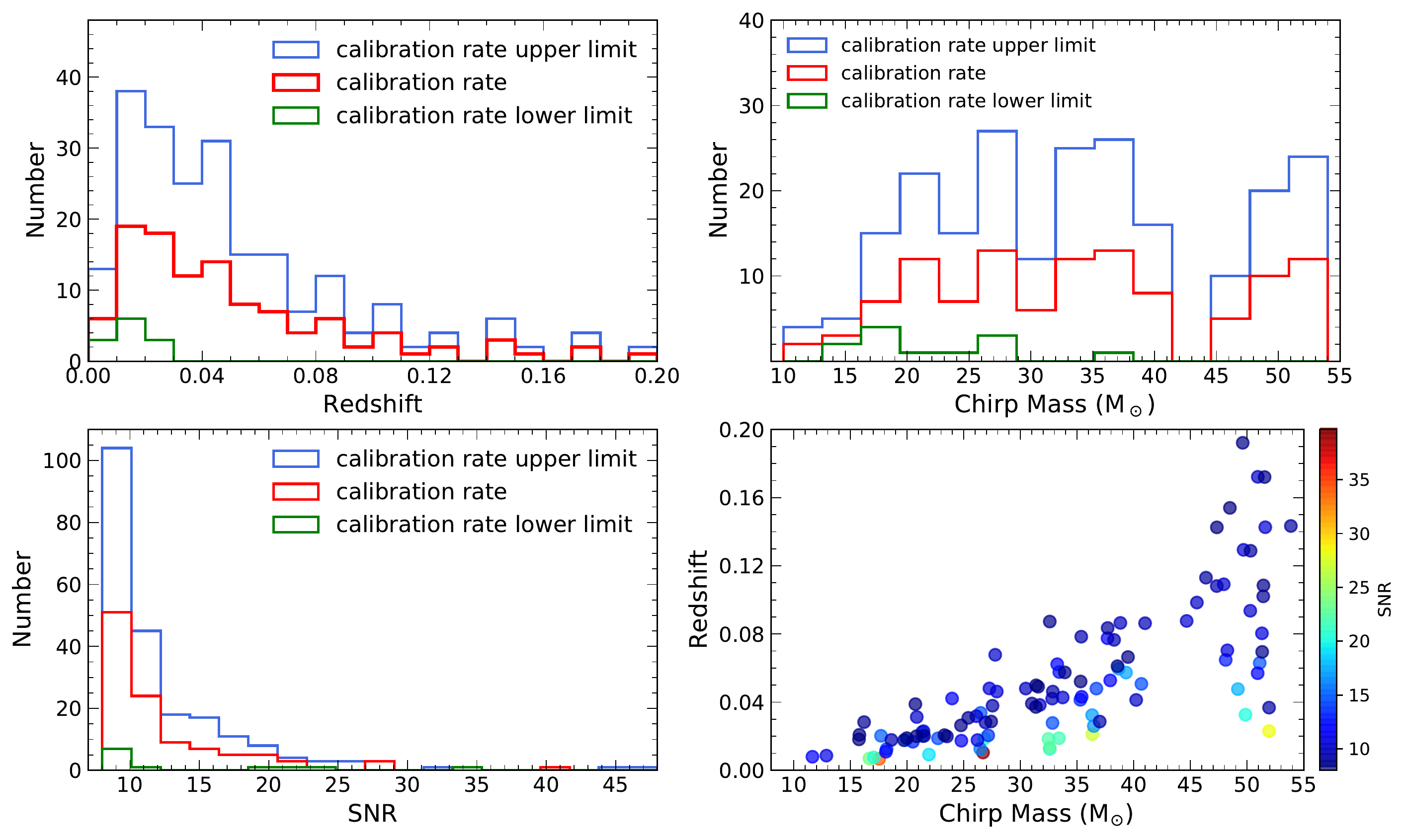}
\caption{ Legend similar to that for Fig.~\ref{fig:f7}, but for ``detectable'' mock BBHs resulting from the  purely dynamical model.
} 
\label{fig:f8}
\end{figure*}

Figures \ref{fig:f7} and \ref{fig:f8} show the distributions of the redshift, chirp mass, and the expected SNR ($>8$  by $5$ years observation of LISA) for those ``detectable'' mock BBHs, resulting from the EMBS dominated model and purely dynamical model, respectively. As seen from these two Figures, almost all ``detectable'' sources are nearby BBHs with redshift $<0.2$ and the closest one has a redshift of $0.005$ and distance of $22$Mpc. In EMBS dominated model, the dynamical channel contributes about half of the detectable sources although its contribution to the total merger rate density is only about a quarter (top two panels of Fig.~\ref{fig:f7}), because of the relatively larger chirp masses and thus larger GW signals of the BBHs produced by the dynamical channel (bottom right panel of Fig.~\ref{fig:f7}). 

\begin{figure}
\begin{centering}
\includegraphics[scale=0.6]{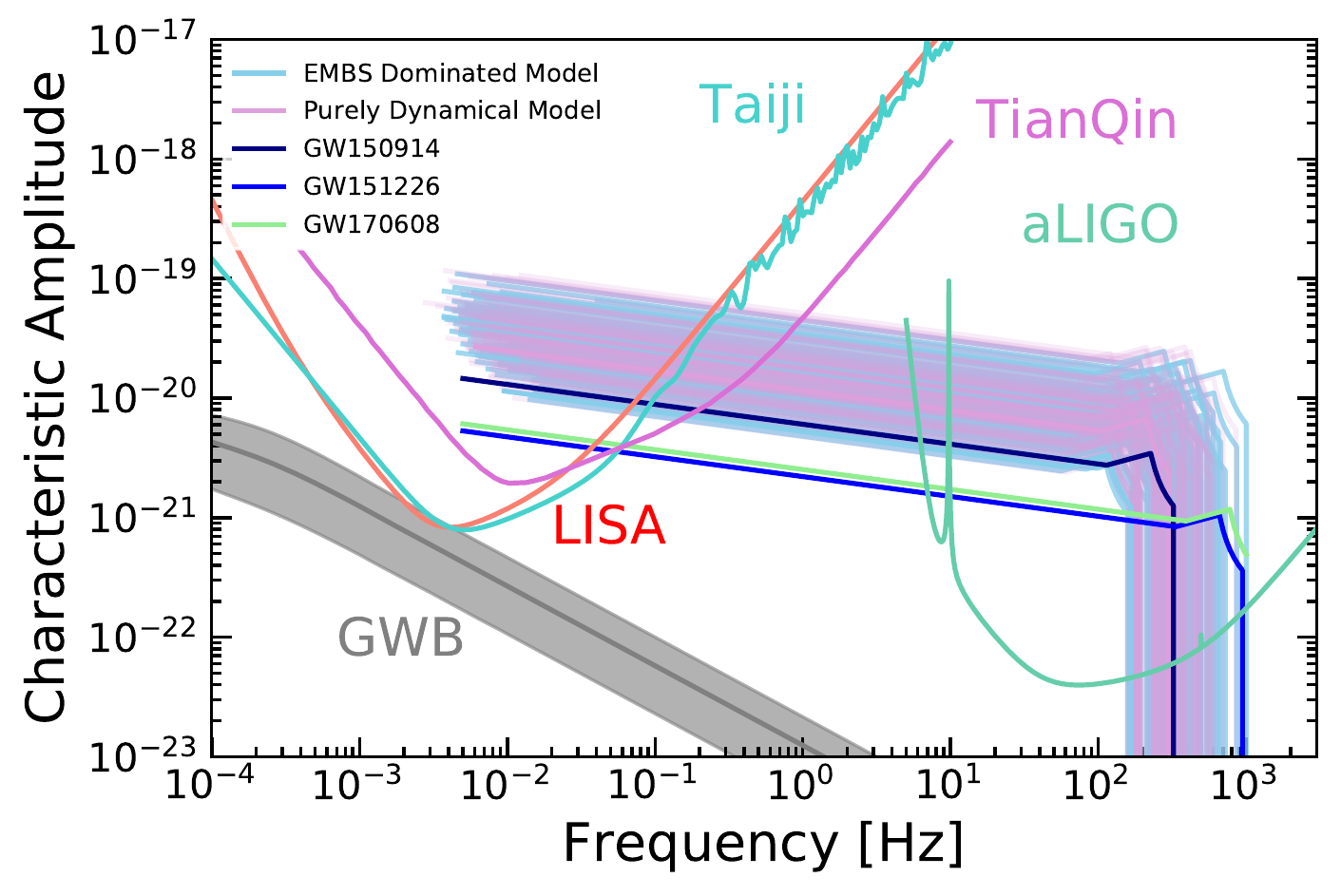}
\caption{The multiband observations of sBBHs from mock samples described in Section \ref{sec:sample} and a few GW sources detected by LIGO O1 and O2 \citep{2019PhRvX...9c1040A}. Light blue lines are the 94 sBBHs from EMBS dominated model with SNR $> 8$ after $5$ years observation of LISA, and violet lines represent 110 sources from purely dynamical model.
GW150914 is shown by black solid line. Red, cyan, violet, blue, and green curves show the sensitivity curves of LISA, Taiji, TianQin, LIGO O2, and LIGO design, respectively. And the grey area shows the confusion of GWB of model {\bf R3:1e1\_d0e2-4\_g9}.
}
\label{fig:f9}
\end{centering}
\end{figure}

We also show the characteristic amplitudes for sources that can be detected by LISA/Taiji/TianQin and by LIGO in Figure \ref{fig:f9} , which demonstrates the possibility of multiband observations of these sources. In the EMBS dominated model, about $94^{+90}_{-89}$ sBBHs can be detected with SNR $> 8$ by LISA after $5$ years observation. Among these sources $17^{+17}_{-16}$  can merge within $5$ years and detected by ground-base GW detectors. In the purely dynamical model, among $110^{+111}_{-98}$ sBBHs that can be observed by LISA in $5$ years, $13^{+13}_{-13}$ sources will merge in $5$ years and can be detected by LIGO. The confusion of GWB is shown as the grey area in Figure \ref{fig:f9}.
We also give the corresponding numbers for Taiji and TianQin in Table \ref{tab:multinum}. We find that about $168^{+165}_{-161}$ sBBHs in the EMBS dominated model can be detected with SNR $> 8$ by Taiji and about $44^{+44}_{-43}$ of them will merge within $5$ years and can be detected by LIGO. For the purely dynamical model, the corresponding numbers are $180^{+185}_{-165}$ and $43^{+44}_{-43}$, respectively. As for TianQin, in the EMBS dominated model about $112^{+111}_{-109}$ sBBHs have SNR $> 8 $ after $5$ years' observation and $100^{+101}_{-99}$ of them will merge within $5$ years and can be detected by LIGO. In the purely dynamical model, the corresponding numbers are $97^{+98}_{-93}$ and $84^{+85}_{-84}$, respectively. Our results on TianQin are consistent with those in the work of \citet{2020PhRvD.101j3027L}. Taiji detects more sBBHs than LISA and TianQin, while TianQin has advantages in detecting multiband sources with relatively shorter lifetimes as it can probe higher frequencies. We also find that $14^{+13}_{-10}$ and $6^{+6}_{-6}$ mock BBHs may be ``detected'' with SNR$>18$ for the purely dynamical model and EMBS dominated model (see the bottom-right panels in Figs.~\ref{fig:f7} and \ref{fig:f8}), respectively, which means they can be ``detected'' with SNR$>8$ by the first year observations of LISA as SNR is roughly proportional to the square root of the observation time. With the first year observations of Taiji and TianQin, $21^{+22}_{-17}$ ($14^{+15}_{-12}$) and $11^{+11}_{-10}$ ($12^{+12}_{-11}$) mock BBHs may be ``detected" for the purely dynamical model (or the EMBS dominated model), respectively.

The fact that the GWB is significantly high than the GW signal of individual sBBHs and BNSs raises significant challenge in the data analysis for extracting individual stellar compact binaries from future LISA/Taiji/TianQin observations \citep[see, e.g.,][]{2018APS..APRH14001B}. We defer further studies on such data analysis to future. Note that a number of recent studies have been investigating whether multi-band observations could be down by combining ground-based GW observatories, such as LIGO/VIRGO/KAGRA, and LISA observations \citep{2016PhRvL.116w1102S, 2019PhRvD..99j3004G, 2019MNRAS.488L..94M}. Furthermore, it may be easier to find the progenitors of some merging stellar compact binaries by digging the LISA (and Taiji/TianQin) archive data if they were detected by ground-based GW observatories with accurate parameter estimation \citep{2019MNRAS.488L..94M}.

\begin{table*}
\centering
\caption{Number of ``Detectable" mock sBBHs by LISA, Taiji and TianQin.}
\label{tab:multinum}
\begin{tabular}{|c|cc|cc|} 	\hline
\multirow{2}{*}{Detector} & \multicolumn{2}{c}{EMBS Dominated Model} & \multicolumn{2}{c}{Purely Dynamical Model} \\ \cline{2-3} \cline{4-5} 
& SNR $> 8$ & SNR $> 8$ and $\tau < 5$ years & SNR $> 8$ &  SNR $> 8$ and $\tau < 5$ years  \\   \hline
LISA & $94^{+90}_{-89}$ & $17^{+17}_{-16}$ & $110^{+111}_{-98}$ & $13^{+13}_{-13}$ \\ \hline
Taiji & $168^{+165}_{-161}$ & $44^{+44}_{-43}$ & $180^{+185}_{-165}$ & $43^{+44}_{-43}$ \\ \hline
TianQin & $112^{+111}_{-109}$ & $100^{+101}_{-99}$ & $97^{+98}_{-93}$ & $84^{+85}_{-84}$ \\ \hline
\end{tabular}
\begin{flushleft}
\footnotesize{Note: the observations time period for LISA (Taiji,TianQin) are set as $5$\,years. First column denotes the GW detectors. Second column shows the number of sBBHs with SNR $> 8$ in EMBS dominated model. Third column shows the number of detectable sBBHs that will merge within $5$ years in EMBS dominated model. Fourth and fifth columns represent those corresponding numbers in purely dynamical model.}
\end{flushleft}
\end{table*}

\section{Discussions}
\label{sec:discussion}

In this paper, we adopt simple semi-analytic models to estimate the cosmic population of sBBHs formed via the EMBS and dynamical channels and BNSs formed via the EMBS channel and further estimate the GWB energy density spectrum.
In principle, these simple models may be improved by the combination of more sophisticated population synthesis models, in which more detailed physical processes involved in the formation of both sBBHs and BNSs can be considered, with cosmological galaxy formation and evolution models, and thus improving the estimates of the merger rates of stellar compact binaries and the distributions of their properties. In the present paper, we did not consider such more sophisticated models partly for the following reasons. There are still large uncertainties in those population synthesis models because many of the physical processes involved in the formation of sBBHs and BNSs are not well understood \citep[e.g.,][]{2016Natur.534..512B}, and the resulting merger rates from different models differ a lot. Especially, most of the BNS formation models lead to a locate merger rate only marginally consistent with the lower bound of the current LIGO/VIRGO constraint \citep[e.g.,][]{2015ApJ...814...58D,2018MNRAS.479.4391M, 2019MNRAS.486.2494G}. If the real BNS merger rate is significantly smaller than the current constraint, the bending of the GWB spectrum at low frequencies would be more evident.

To investigate the effect of highly eccentric sBBHs formed via the dynamical channel or other channels on the GWB spectrum and for demonstration purpose, we simply assume different  
eccentricity distributions at an early stage for BNSs and (EMBS originated and dynamically originated) sBBHs. Our results show that the GWB spectrum is bent significantly if the contribution from highly eccentric sBBHs originated from the dynamical channel to the GWB is significant. In future, one could also directly obtain the eccentricity distribution of dynamically originated sBBHs from detailed dynamical modelling of dense stellar systems over the cosmic time by adopting more complicated models. We adopt a simple assumption on the eccentricity distribution, Gaussian-like distribution, of GW sources. However, the real eccentricity distribution for sBBHs and DNSs are highly uncertain \citep{2020ApJ...892L...9A}. Results from different models shown in Table~\ref{tab:model} suggest that different eccentricity distributions lead to different spectrum indices of GWB at the low-frequency end, which may be used to distinguish/constrain models.

We also did not consider the contributions to the GWB by sBBHs formed via the AGN/MBH-assisted channel, the PBH channel, and those sBBH mergers induced by the Lidov-Kozai mechanisms in triple systems because of large uncertainties in their merger rate estimates. In general, those sBBHs formed via these several mechanisms may also have larger eccentricities, thus they can have a similar effect to the bend of the GWB spectrum as that of the dynamically originated sBBHs. In future, one may consider all those mechanisms together to form a comprehensive model for estimating the cosmic sBBH population and their contribution to the GWB. Or in the other way, one may generate a population synthesis model for the ``detected" GWB spectrum with additional information from individual sources detected by the ground-based GW observatories to constrain the contributions from different channels to the cosmic population of sBBHs, etc.  

Note here that the mergers of neutron star-black hole binaries (NSBHs), are not considered in the present work, simply because there is no available constraint on its merger rate, yet, though there are some candidates already discovered by the O3 observations of LIGO/VIRGO (http://gracedb.ligo.org). Assuming the local merger rate density for NSBHs is $\sim 40\gpcyr$ \citep[see][]{2018MNRAS.479.4391M} and the typical chirp mass is $\sim 3.0 M_{\odot}$ (for typical NSBHs $1.4M_{\odot} + 10M_{\odot}$), we find that their contribution to the GWB is roughly $\sim 7\%$ of that from sBBHs and BNSs listed in Table \ref{tab:model}. However, if the merger rate density of BHNS is significantly higher than $40\gpcyr$, their contribution to the GWB could also be significant.

The GWB may act as a confusion to the detection of individual GW events. We have shown that this confusion may lead an SNR decrease when detecting individual stellar compact binaries though it is not substantial. Since the GWB is expected to be measured by the space GW detectors with high SNR($\gtrsim 280$) and may be well modelled, the SNR decrease due to such a GWB confusion can be reduced to the minimal. For other GW sources, such as mergers of binaries of massive/intermediate-mass black holes and extreme-mass-ratio-inspirals (EMRIs), similarly, the confusion from the GWB due to stellar compact binaries will not lead to any significant decrease to their detection SNR. Note here in our estimates of the GWB only the cosmic stellar compact binaries are considered, but mergers of intermediate-mass black holes (IMBHs) and EMRIs may also contribute to the GWB in the low frequency band ($10^{-4}-1$\,Hz). The rates of IMBH mergers and EMRI event are highly uncertain, dependent on the poorly known seed black hole formation mechanisms and detailed structures of galactic nuclei, etc., which hind robust estimates of their contributions to the GWB. 

\section{Conclusions}
\label{sec:conc}

In this paper, we investigate the GWB contributed by stellar compact binaries at frequencies from $\sim 10^{-4} -1000$\,Hz. We consider the contributions both from cosmic populations of sBBHs formed via evolution of massive binary stars (the EMBS channel) and dynamical interactions of compact (binary) stars in dense stellar systems (the dynamical channel) and cosmic BNSs formed via the EMBS channel. By investigating various simple models for the formation of sBBHs and BNSs and their property distributions, especially, the eccentricity distribution resulting from the dynamical channel, we find that the GWB spectrum in the low frequency band ($10^{-4}-1$\,Hz; the band for space GW detectors like LISA/Taiji/TianQin) may not be the unique power-law with a slope of $2/3$ in the whole frequency range, but bent significantly at lower frequencies due to the high eccentricities of sBBHs formed via the dynamical channel and can be fitted by a double-power law with a slope $\simeq 2/3$ at high frequency part but a slope substantially larger than $2/3$ at low frequency part. The significance of such a bend depending on the significance of the contribution from sBBHs formed via the dynamical channel to the total cosmic population of sBBHs. If this contribution is less than a fraction of $\sim 25\%$, the GW spectrum is only slightly bent, and the difference between the two slopes for the double-power fitting to the GWB is small ($\la 0.2$). However, the bending is quite significant if the dynamical originated sBBHs dominate the sBBH cosmic population and the difference between the two slopes can be large ($\ga 0.4-0.5$). The turnover frequency of the double power-law GWB spectrum is $\sim 1\times 10^{-3}$\,Hz if the dynamical originated sBBHs have high eccentricities ($\gtrsim 0.9-0.7$) at $f\sim 10^{-4}-10^{-3}$\,Hz, and it may move to a higher frequency if these sBBHs can have high eccentricities at higher frequencies.

Our results show that the GWB at the low-frequency band ($10^{-4}-1$\,Hz) can be detected by LISA/Taiji/TianQin over a mission time of $5$\,years with SNR $\gtrsim$ $274/255/21$, and it can be detected with SNR $\gtrsim$ $18/17/1.5$ with the first week observations of LISA/Taiji/TianQin. This suggests that the GWB from stellar compact binaries may be  the first GW signal to be revealed by the space GW detectors. The reason is that the event rates of other main sources, such as the mergers of massive binary black holes or IMBHs, are estimated to be mostly less than a few to a few tens per year \citep[e.g.,][]{2005ApJ...623...23S, 2018ApJ...867..119F, 2019MNRAS.488.4370F, ChenYuLu20}. Although the rate for EMRIs could be in the range from a few to thousands per year \citep[][]{2017PhRvD..95j3012B}, but such estimates have large uncertainties. Detailed analyses suggest that EMRIs may be or may be not detected in the first week observation of LISA/Taiji/TianQin \citep[e.g.,][]{2017JPhCS.840a2021G}. This is quite different from the case for the high-frequency GWB ($1-1000$\,Hz), for which many individual sources have already been detected by LIGO/VIRGO while the GWB is still waiting to be detected in about $6$ years \citep{2018PhRvL.120i1101A} (assuming the full sensitivity of LIGO achieved in 2023 and after 3 years of LIGO observation).

We estimate the GWB in the time domain by summing up GW signals from numerous mock cosmic sBBHs and BNSs generated from our models. We find that the stochastic GWB in the time domain has an amplitude of $\sim 10^{-21}$, which is around two orders of magnitude larger than the source with the largest SNR. The confusion from the GWB does have some effect on the SNR estimate for individual stellar compact binaries, however, its effect is not so significant. Furthermore, the GWB may be well detected and modelled when searching individual stellar compact binaries and thus the effect may be reduced to be the minimal.

We also investigate the detection of individual stellar compact binaries by space GW detectors and find that within the mission time of $5$ years LISA/Taiji/TianQin may detect about $\sim5$-$221$/$7$-$365$/$3$-$223$ (or $\sim16$-$782$/$27$-$1645$/$6$-$1479$) sBBHs with SNR$\gtrsim 8$ (or $5$) individual sBBHs, but none of the cosmic BNSs (not the Milky Way BNSs) can be detected with high SNRs (e.g., $\gtrsim 3$). With the first year observations of LISA/Taiji/TianQin, the number of that sBBHs may be ``detected'' with SNR$\gtrsim 8$ is expected to be $\sim 0$-$27$/$2$-$43$/$1$-$24$. If the mission time is extended to $10$ years, then the number of sBBHs that can be detected increase to $\sim14$-$539$/$19$-$936$/$7$-$729$ (or $\sim39$-$3229$/$59$-$7975$/$12$-$4879$) with SNR$\gtrsim 8$ (or $\gtrsim 5$). Among these detectable sources (with SNR$\gtrsim$ 8) during $5$ years mission of LISA/Taiji/TianQin, about $\sim0$-$34$/$0$-$88$/$0$-$201$ (or $\sim0$-$49$/$1$-$143$/$0$-$211$) sBBHs can merge within $5$ (or $10$) years. 

\section*{Acknowledgements}
This work is partly supported by the National Natural Science Foundation of China (Grant No. 11690024, 11873056, 11991052), the Strategic Priority Program of the Chinese Academy of Sciences (Grant No. XDB 23040100), and the National Key Program for Science and Technology Research and Development (Grant No. 2016YFA0400704).

\section{Data Availability}

The data underlying this article are available in the article.









\end{document}